\documentstyle[aps,prd,epsfig]{revtex}

\begin{document}

\draft

\title{Effective action and motion of a cosmic string}

\author{Malcolm Anderson\cite{mand}}
\address{Department of Mathematics, Edith Cowan University \protect\\
         2 Bradford Street, Mount Lawley, WA 6050, Australia}

\author{Filipe Bonjour\cite{fbon} and Ruth Gregory\cite{rgre}}
\address{Centre for Particle Theory \protect\\
         University of Durham, Durham, DH1 3LE, U.K.}

\author{John Stewart\cite{jste}}
\address{Department of Applied Mathematics and Theoretical Physics \protect\\
       Silver Street, Cambridge, CB3 9EW, U.K.}

\date{\today}

\maketitle

\begin{abstract}
  We examine the leading order corrections to the Nambu effective action
  for the motion of a cosmic string, which appear at fourth order in the
  ratio of the width to radius of curvature of the string.  We determine
  the numerical coefficients of these extrinsic curvature corrections, and
  derive the equations of motion of the worldsheet. Using these equations,
  we calculate the corrections to the motion of a collapsing loop, a travelling
  wave, and a helical breather. From the numerical coefficients
  we have calculated, we discuss whether the string motion
  can be labelled as `rigid' or `antirigid,' and hence whether cusp or
  kink formation might be suppressed or enhanced.
\end{abstract}

\pacs{11.27.+d, 11.10.Lm, 98.80.-k \\[3mm]
      \hfill DTP/97/5 \hfill DAMTP R-97/27
      \hfill hep-ph/9707324}

\section{Introduction}

The study of topological or vacuum defects
is of importance in many areas of contemporary physics.
In high energy physics, a defect will generically occur
during a symmetry breaking process where different parts of a medium
choose different vacuum energy configurations, and the
non-compatibility of these different vacua forces a sheet, line, or point of
energy where these non-compatible vacua meet. The relevant
vacuum order parameter then becomes indeterminate---this is the defect.
A defect may be topological~\cite{VS}, in that it is the topology of the
vacuum that simultaneously allows formation, and prevents dissipation, of
these objects---but other types of defect are also possible. For instance,
a defect may be stable dynamically (i.e.~classically, due to
energy considerations) but not topologically, as it happens for semilocal
\cite{VA} or electroweak \cite{V}
defects. A defect can even be `topological'
and unstable, as in the case of textures~\cite{T}, but nonetheless of physical
importance.

In cosmology, there has been much speculation that topological defects
(stable or otherwise) might have played an important role in structure
formation~\cite{defect}. In general, there are two
main concerns when considering the cosmological effects of topological defects:
their gravitational effects and their dynamics. Any theory concerning galaxy
formation must be able to allow or constrain the
presence of strongly self-gravitating objects. But the dynamics of the
defects are in fact even more important, for it is the dynamics that
determine to a large extent the shapes of the gravitating defects. 
For instance, if cosmic strings did not intercommute, any network would
rapidly become tangled and would not obey a scaling law; such a 
configuration would be in conflict with the universe we see around us
today. Even with intercommutation~\cite{Int}, strings which are strongly rigid
and hence straight will have different gravitational effects
to those that are very crinkly~\cite{SG,Cri}.

The dynamical behaviour of a defect is generally assumed to be
approximated by an effective action, a description which models
the rather large numbers of degrees of freedom of the full field
theory by the smaller number of degrees of freedom based on the 
position of the core of the defect.
Attempts to derive effective actions or equations of motion for topological
defects have commonly focussed on the strong coupling limit, meaning that of
large values of the coupling coefficient $\lambda$ of the relevant Higgs
field. In this limit, the defect becomes infinitesimally thin and
effectively decouples from the other (infinitely massive) particles in the
field theory. 
The study of the effective motion of topological defects has
been extended~\cite{MT,GG,SOA} away from the limit
$\lambda\rightarrow \infty$ to
cases for which the thickness is small but not exactly zero. The resultant
effective action generically contains a `zero-thickness' term proportional
to the area of the defect~\cite{NO,F}, 
and extrinsic curvature terms which appear at
quadratic order in the thickness~\cite{MT,GG,SOA}. 
The exact form of these second order terms is dependent on whether an
`off-shell' or `on-shell' approach has been used to derive the
effective action as was explained using the example of the domain wall
in~\cite{CG}, and one finds that in a self-consistent order by order
solution of the equations of motion, the only quadratic correction appearing
is due to the geometry of the defect worldsurface, and is proportional
to its intrinsic Ricci curvature~\cite {CG}. For the domain wall, this 
term gives corrections to the motion, however for the string this term is
a topological invariant---proportional to
the Euler character of the worldsheet---and hence does not give any 
correction to the Nambu equations of motion.

The leading order corrections to the motion of a cosmic string were shown by
one of us, \cite{A}, to appear at quartic order in the string width. 
In this paper, a systematic expansion of the geometry and field
equations clarified an earlier discrepancy concerning second order
`twist' terms~\cite{MT}, and derived the fourth order action of the string,
which is
\begin{equation}
S = - \mu \int d^2 \sigma \sqrt{-\gamma} \bigg[ 1 - r_s^2
      {\alpha_1\over\mu} {\cal R} + r_s^4 \bigg( {\alpha_2\over\mu}
      {\cal R}^2 + {\alpha_3\over\mu}{\cal K}_{_i\mu\nu}{\cal K}_{_j}^{\mu\nu}
      {\cal K}_{_i\lambda\rho} {\cal K}_{_j}^{\lambda\rho} \bigg) \bigg],
      \label{anderson}
\end{equation}
where ${\cal R}$, ${\cal K}_{_i\mu\nu}$ are the Ricci and extrinsic
curvatures of the worldsheet (to be defined in the next section), and
the $\alpha_i$ are numerical coefficients depending on the field theory
modelling the vortex (to be defined in section~\ref{sec:correct} for
the Abelian-Higgs model). Unlike the domain wall, whose background solution,
and extrinsic
curvature corrections can be expressed analytically~\cite{CG}, the
prototypical cosmic string solution, the Nielsen--Olesen vortex~\cite{NO},
does not have a closed analytic form, hence these coefficients must be
evaluated numerically.

As soon as one calculates corrections to the Nambu action, it becomes
of general concern whether or not these corrections cause the motion of the 
defect to be `rigid' or `antirigid.' The implications of extrinsic curvature
terms for string motion have been well-explored~\cite{GG,CGZ,Po,Kl,Car}, 
however, it is not always
easy to decide {\it a priori\/} (especially for the fourth order terms) 
whether the strings will be rigid or not, or indeed what one means by
rigid~\cite{Car}.

In this paper we address these issues. 
We determine the numerical values of the $\alpha_i$ for 
the Abelian-Higgs model by solving the perturbed field equations 
for the non-flat worldsheet.
We then derive the fourth order equations of motion from Anderson's 
action and calculate 
corrections to three `test case' trajectories: 
the circular loop, the travelling wave, and the helical breather.
These three solutions display different characteristics which 
should be mirrored in the corrections to the Nambu motion if 
the rigidity of the string is to be determined. 
The loop collapses to a point~\cite{KT}, and rigidity would be indicated 
by a retardation of this collapse. A travelling wave on the 
other hand has been shown to be an exact solution of the full 
field theory~\cite{TW}, and should not exhibit any corrections if our 
approach is to be trusted.
The helical breather (see e.g.~\cite{HB}) is a time dependent solution
which is never singular. Rigidity for this trajectory is more subtle, since the
helical breather is never singular, however, we could call the string rigid
if the tendency of the correction is to lower the magnitude of the scalar
curvature of the
worldsheet. As we will see, it is rather difficult to give an intuitive
criterion for rigidity.

The layout of the paper is as follows. In the next section we review
the formalism for the derivation of the effective action. In section~%
\ref{sec:correct} we rederive Anderson's action, and present new numerical
results evaluating the coefficients appearing in the action. In section~%
\ref{sec:eom} we derive the equations of motion of the fourth order string,
and in section~\ref{sec:traj} we calculate corrections to three test
trajectories. In section~\ref{sec:rigid} we discuss the question of rigidity
and conclude.

\section{Deriving the effective action}

In this section we review the formalism required for the derivation
of the string effective action. This formalism is largely based on the
contents of~\cite{GG,A}. 
The problem of building the effective action is to reduce some 
four-dimensional field theoretic action integral
\begin{equation} \label{actfour}
  S = - \int d^4 x \sqrt{-g} {\cal L}
\end{equation}
to some two-dimensional worldsheet integral
\begin{equation} \label{actwo}
S_{\rm eff} = - \int d^2 \sigma \sqrt{-\gamma} {\cal L}_{\rm eff}.
\end{equation}
We follow the canonical approach by setting up a coordinate 
system based on the worldsheet, and expanding the action and 
equations of motion around this worldsheet. By `expand' we mean 
that we do not expect in one go to be able to solve the full 
equations of motion and integrate out (otherwise why find an 
effective action) but that we will be able to express the 
equations of motion for the system in terms of an expansion in 
$r_s\kappa$, where $r_s$ is the radius of the string, and $\kappa$ a 
typical scale of the extrinsic curvature of the worldsheet.
We then solve the equations of motion in the directions perpendicular 
to the worldsheet, and integrate out the four-dimensional
action~(\ref{actfour}) over these degrees
of freedom to get an action of the form~(\ref{actwo}).

We will look at the particular case of a U(1) local string in 
flat spacetime (signature $+,-,-,-$). This is a vortex solution of the 
Abelian-Higgs model
\begin{equation} \label{abhiggs}
  {\cal L} = \left( {\cal D}_\mu \varphi \right) ^\dagger
  \left( {\cal D}^\mu \varphi \right)
  - {1\over 4} {\tilde F}^{\mu\nu}
  {\tilde F}_{\mu\nu}
  - {\lambda \over 4} (|\varphi|^2 - \eta^2)^2,
\end{equation}
where ${\cal D}_\mu = \nabla_\mu + ieA_\mu$ is the usual gauge 
covariant derivative, and ${\tilde F}_{\mu\nu}$ the field 
strength associated with $A_\mu$. We rewrite the field content of 
this model in the usual way:
\begin{mathletters} \label{changevar}
  \begin{eqnarray}
    \varphi (x^\alpha) &=& \eta \, X(x^\alpha)e^{i\chi(x^\alpha)}  \\
    A_\mu(x^\alpha)    &=& \frac1e \, \bigg[ P_\mu (x^\alpha) - \nabla_\mu
    \chi (x^\alpha) \bigg],
  \end{eqnarray}
\end{mathletters}
so that the vacuum is characterised by $X=1$. In terms of these 
new variables, the Lagrangian becomes
\begin{equation}\label{newlag}
  {\cal L} = \eta^2 (\nabla_\mu X)^2 + \eta^2 P_\mu^2 X^2  - \frac1{4e^2} 
  F_{\mu\nu}^2 - \frac{\lambda \eta^4}4 (X^2-1)^2,
\end{equation}
where $F_{\mu\nu}$ is now the field strength of $P_\nu$. By 
making a compensating gauge transformation in $A_\mu$, we have 
absorbed the unphysical gauge degree of freedom of the theory. For 
future reference, here are the equations of motion for the new 
fields:
\begin{mathletters} \label{eomxp}
  \begin{eqnarray}
    \frac1{\sqrt{-g}} \, \partial_\mu \sqrt{-g} \, \partial^\mu X &=&
      P_\mu P^\mu X - \frac{\lambda \eta^2}2 X(X^2-1) \\
    \frac1{\sqrt{-g}} \, \partial_\mu \sqrt{-g} \, F^{\mu\nu} &=&
      - 2 e^2 \eta^2 X^2 P^\nu.
  \end{eqnarray}
\end{mathletters}
These equations are known to admit vortex solutions, the best known being
that due to Nielsen and Olesen~\cite{NO}---a solution corresponding
to an infinite straight static
string. A vortex solution is characterised by the fact that the scalar 
field phase has a non-zero winding number around an axis 
corresponding to the zeroes of the Higgs field (X=0). If we choose 
the string axis to be aligned with the $z$-axis, then a gauge 
can  be chosen in which the vortex solution takes the form
\begin{equation}\label{nov}
  \varphi = \eta \, X_{\rm NO}(\sqrt{\lambda}\eta\rho) e^{i\theta}, \qquad
    A_\mu = {1\over e} \left [ P_{\rm NO} (\sqrt{\lambda}\eta\rho) -1 \right ]
    \nabla_\mu \theta
\end{equation}
in cylindrical polar coordinates, where $X_{\rm NO}$ and $P_{\rm NO}$
satisfy the (generically) second order coupled differential equations:
\begin{mathletters} \label{no}
  \begin{eqnarray}
  -X'' -{X'\over r} + {XP^2\over r^2} +{1 \over 2}X(X^2-1) &=& 0 \\
  -P'' + {P'\over r} + \beta P X^2 &=& 0
  \end{eqnarray}
\end{mathletters}
and we have introduced the Bogomol'nyi parameter $\beta = 2e^2/\lambda$.
$\beta \to 0$ is the global string, and $\beta = 1$ is the critical
(supersymmetric) string for which the equations of motion factor to two
coupled first order equations:
\begin{mathletters}
  \begin{eqnarray}
    r X' &=& X P, \\
    P'   &=& \frac12 r \big( X^2 - 1 \big).
  \end{eqnarray}
\end{mathletters}

Intuitively we expect the solution for a general worldsheet to be 
very close to the Nielsen--Olesen solution in suitable 
coordinates, and that we might be able to expand the fields, and
therefore the action, around this approximate solution. Let us now 
try to make this idea more concrete.
 
First we choose a worldsheet-based coordinate system. (Note this 
will only be valid within the radii of curvature of the 
worldsheet.) We start by coordinatising the worldsheet by 
$\sigma$ and $\tau$ as usual. Since we are in flat space, at 
each point on the worldsheet we have a well defined orthogonal 
flat two-plane, and such orthogonal planes do not intersect within 
the radii of curvature of the worldsheet. Thus we can thicken the 
worldsheet to a worldblanket by defining $\sigma$ and $\tau$ to 
be constant on such orthogonal planes. Since the worldsheet 
$\cal W$ has codimension 2, it has two associated families of 
unit normals $\{n_{_i}^\mu\}$. 
We choose a cartesian parametrisation 
of the orthogonal planes such that $\left ( {\partial \over 
\partial \xi^i} \right ) ^\mu _{\xi^i = 0}= n_{_i}^\mu$. 
Thus 
$\{\tau,\sigma,\xi^i\}$ define a coordinate system in the vicinity 
of the worldsheet. In general it will not be possible to choose 
this system to be globally orthogonal, however, we will assume 
that the normal fields have been chosen so that the departure 
from orthogonality is minimal, i.e. of order of the extrinsic 
curvature of the worldsheet.

Because we no longer have a flat coordinate system, the 
connection will no longer be trivial, although the curvature 
components should still vanish. In order to examine the form of 
these relations, we use a Gauss--Codazzi approach~\cite{Spivak} (see
also~\cite{Carter} for more of a physicist's approach). For shorthand 
we denote by $\{\sigma^A\}$ ($A=0,1$) the coordinates $\{ \tau, 
\sigma \}$, the worldsheet by $\cal W$ and the Minkowski spacetime by $\cal M$.
 
The {\it first fundamental form\/} $h_{\mu\nu}$ of $\cal W$ is defined 
as
\begin{equation}\label{fff}
  h_{\mu\nu} = g_{\mu\nu} + n_{_i\mu} \, n_{_i\nu}.
\end{equation}
The tensor $h_{\mu\nu}$ acts as a projector onto $\cal W$, but is still 
a tensor in the 4-dimensional spacetime. For the metric {\it 
intrinsic\/} to the worldsheet, one uses the familiar
\begin{equation}\label{intrinsg}
  \gamma_{AB} = {\partial X^\mu \over \partial \sigma^A}
  {\partial X_\mu \over \partial\sigma^B},
\end{equation}
where $X^\mu(\sigma^A)$ are the spacetime coordinates of the 
submanifold $\cal W$. This gives an interpretation of $\cal W$ as both a 
worldsheet in spacetime and a two-dimensional manifold with its own 
intrinsic geometry. One can define the intrinsic curvature of 
$\cal W$ in the standard way, and the extrinsic curvatures, or
{\it second fundamental forms}, by
\begin{equation}\label{sff}
  {\cal K}_{_i\mu\nu} = h_{(\mu}^\rho h_{\nu)}^\sigma
  \nabla_\rho n_{_i\sigma}.
\end{equation}
These extrinsic curvatures ${\cal K}_{_i\mu\nu}$ measure how $\cal W$ curves
in $\cal M$.
Since the codimension of the worldsheet is greater than one, we 
also have a non-trivial {\it normal fundamental form}
\begin{equation}\label{nff}
  {\tilde\omega}_\mu = \frac12 \varepsilon_{ij} n_{_j\nu} \nabla_\mu
  n_{_i}^\nu,
\end{equation}
where $\varepsilon_{ij}$ is the alternating symbol on two 
indices.
This represents how near to orthogonality the worldblanket 
coordinate system is; it measures the rotation of the 
$n_{_i}^\mu$ in their own planes as one moves around the worldsheet.
Note that, despite lying tangent to $\cal W$, $\tilde{\omega}_\mu$ is a
gauge dependent object, depending on the choice of the normal 
fields---an SO(2) gauge group. 
In fact, $\tilde{\omega}_\mu$ is the connection on the normal 
bundle of $\cal W$.

Now, we may write
\begin{equation}\label{dern}
  \nabla_{(\mu} n_{_i\nu)} = {\cal K}_{_i \mu\nu}
  - \varepsilon_{ij} {\tilde\omega}_{(\mu}
  n_{_j\nu)},
\end{equation}
where the symmetrization is understood to be only acting on indices
of the same type, so that the `$j$' above does not participate in any
symmetrization.
This now indicates an alternative definition of the second fundamental forms:
\begin{equation}
  {\cal K}_{_i\mu\nu} = \frac12 {\cal L} _i g_{\mu\nu}
    + \varepsilon_{ij} {\tilde\omega}_{(\mu}n_{_j\nu)} =
    \frac12 {\cal L}_i h_{\mu\nu}. \label{altk}
\end{equation}
where ${\cal L}_i$ denotes the Lie derivative with respect to 
$n_{_i}^\mu$. We may now use the Riemann identity in flat space 
to derive the following relation for the extrinsic curvatures:
\begin{equation}\label{riem}
  R_{\sigma(\mu\nu)\rho} n_{_i}^\sigma n_{_j}^\rho
  =  {\cal L}_j {\cal K}_{_i\mu\nu}
  - {\cal K}_{_i\rho(\mu} {\cal K}_{_j\nu)}^\rho = 0.
\end{equation}
Therefore, contracting with $g^{\mu\nu}$ gives
\begin{equation}\label{liek}
  {\cal L}_j {\cal K}_{_i} = - {\cal K}_{_i\mu\nu} 
  {\cal K}_{_j}^{\mu\nu}
\end{equation}
In addition, note that 
\begin{equation} \label{liebeta}
  {\cal L}_i {\tilde\omega}_\mu = n_{_i}^\nu \nabla_\nu {\tilde\omega}_\mu
  + {\tilde \omega}_\nu \nabla_\mu n_{_i}^\nu = {\frac12} n_{_i}^\nu
  \nabla_\nu \left (\varepsilon_{jk}n_{_k\sigma}\nabla_\mu n_{_j}^\sigma
  \right ) + {\tilde \omega}_\nu \nabla_\mu n_{_i}^\nu = 0
\end{equation}
and
\begin{equation}\label{lien}
  {\cal L}_i n_{_j\nu} = n_{_j\mu} \nabla_\nu n_{_i}^\mu =
  {\tilde\omega}_\nu \varepsilon_{ij}.
\end{equation}

The system of equations~(\ref{altk}--\ref{lien}) give the `equations of 
motion' 
for the geometry of the system, which together with~(\ref{eomxp}) form the 
full equations of motion for the string.

In order to extract a low energy effective action, we will need 
to expand these quantities and equations of motion in terms of the
thickness of the string, $r_s$. To do this systematically we define the
dimensionless parameter $\epsilon$ by
\begin{equation}\label{epsil}
  \epsilon = \frac{\kappa}{{\sqrt \lambda}\eta} \propto \kappa r_s
\end{equation}
where $\kappa^{-1}$ represents a typical radius of curvature of the worldsheet. 
We then define the {\it zero thickness limit\/} to mean $\epsilon \to 0$ 
with $\mu \propto \eta^2$, the energy per unit length of the string,
fixed. Note that this may not necessarily mean that the string width
{\it is\/} zero, since instead the extrinsic curvature could be zero
(a flat worldsheet). However, if $\kappa \neq 0$, this limit
corresponds to the conventional $\lambda\to\infty$ limit.

We now redefine coordinates so that the worldsheet has thickness
and curvature of order unity by setting
\begin{mathletters} \label{rescale}
  \begin{eqnarray}
    x^i &=& \xi^i /r_s \label{rescale_a} \\
    s^A &=& \sigma^A \kappa \label{rescale_b}
  \end{eqnarray}
\end{mathletters}
and redefining for consistency the following variables
\begin{mathletters} \label{redefine}
  \begin{eqnarray}
    K_{_i\mu\nu} &=& {\cal K}_{_i\mu\nu}/\kappa \label{redefine_a} \\
    \omega_\mu   &=& {\tilde\omega}_\mu /\kappa \label{redefine_b} \\
    {\hat P}_\mu &=& r_s P_\mu. \label{redefine_c}
  \end{eqnarray}
\end{mathletters}

The worldsheet geometry equations~(\ref{altk}--\ref{liebeta}) become
\begin{mathletters} \label{unitgeom}
  \begin{eqnarray}
    {\cal L}_i h_{\mu\nu} &=& 2\epsilon K_{_i\mu\nu}  \\
    {\cal L}_i g_{\mu\nu} &=& 2\epsilon [ K_{_i\mu\nu}
      - \varepsilon_{ij} \omega_{(\mu} n_{_j\nu)}] \\
    {\cal L}_j K_{_i\mu\nu} &=& \epsilon K_{_i\sigma(\mu} K_{_j\nu)}^\sigma \\
    {\cal L}_j K_{_i} &=& - \epsilon K_{_i\mu\nu}K_{_j}^{\mu\nu} \\
    {\cal L}_j \omega_\mu &=& 0,
  \end{eqnarray}
\end{mathletters}
and the field equations~(\ref{eomxp}) become:
\begin{eqnarray}
 {1\over \sqrt{-g}} && \bigg\{ \partial_i \bigg[ \sqrt{-g} \big( g^{ij}
   \partial_jX + \epsilon g^{iA}\partial_AX \big) + \epsilon \, \partial_A
   \big( \sqrt{-g}(g^{iA}\partial_iX + \epsilon g^{AB}\partial_B X) \big)
   \bigg] \bigg\} \nonumber \\
 && = {\hat P}_\mu {\hat P}_\nu g^{\mu\nu}X - {\frac12} X(X^2-1) \label{xorder}
\end{eqnarray}
\begin{eqnarray}
  {1\over \sqrt{-g}} && \bbox{\Bigg(} \partial_i \bigg\{ \sqrt{-g} \bigg[
    g^{ik}g^{jl}{\hat F}_{kl}+ (g^{iA}g^{jl}-g^{il}g^{jA})(\epsilon\partial_A
    {\hat P}_l - \partial_l {\hat P}_A) + \epsilon
    g^{iA}g^{jB} {\hat F}_{AB} \bigg] \bigg\} + \nonumber \\
  && \epsilon \, \partial_A \bigg\{ \sqrt{-g} \bigg[ g^{Ak}g^{jl}
    {\hat F}_{kl} + (g^{Ak}g^{jB} - g^{AB}g^{jk})
    (\partial_k {\hat P}_B - \epsilon \partial_B
    {\hat P}_k) + \epsilon g^{AB} g^{jC} {\hat F}_{BC} \bigg] \bigg\}
    \bbox{\Bigg)} \nonumber \\
  && = -\beta X^2 (g^{ij} {\hat P}_i + g^{jA}{\hat P}_A) \label{piorder}
\end{eqnarray}
\begin{eqnarray}
  {1\over \sqrt{-g}} && \bbox{\Biggl(} \partial_i \bigg\{ \sqrt{-g} \bigg[
    g^{ik}g^{Al}{\hat F}_{kl} + (g^{iB}g^{Al}-g^{il}g^{AB})
    (\epsilon\partial_B
    {\hat P}_l - \partial_l {\hat P}_B) + \epsilon
    g^{iB}g^{AC} {\hat F}_{BC} \bigg] \bigg\} + \nonumber \\
  && \epsilon \, \partial_B \bigg\{ \sqrt{-g} \bigg[
    g^{Bk}g^{Al} {\hat F}_{kl} +
    (g^{BC}g^{Al} - g^{Bl}g^{AC})(\epsilon\partial_C{\hat P}_l
    -\partial_l{\hat P}_C) +  \epsilon g^{BC}g^{AD}
    {\hat F}_{CD} \bigg] \bigg\} \bbox{\Biggr)} \nonumber \\
  && = -\beta X^2 (g^{jA}{\hat P}_j + g^{AB}{\hat P}_B). \label{paorder}
\end{eqnarray}
Here, $\hat{F}_{ij}$ is the tensor $F_{ij}$ defined in terms of the rescaled
variables. Finally, we also need the explicit expansion of the Lagrangian
\begin{eqnarray}
  {\cal L} = \lambda \eta^4 && \bbox{\Biggl(} \partial_i X \partial_j X g^{ij}
    + 2\epsilon \partial_i X \partial_A X g^{iA}
    + \epsilon^2 \partial_A X \partial_B X g^{AB}
    + X^2 {\hat P}_\mu {\hat P}_\nu g^{\mu\nu} - \nonumber \\
  && {1\over 2\beta} \bigg\{
    {\hat F}_{ij} \bigg[ {\hat F}_{kl} g^{ik} g^{jl}
    + 4 ( \partial_k {\hat P}_A - \epsilon \partial_A {\hat P}_k) g^{ik} g^{jA}
    + 2\epsilon {\hat F}_{AB} g^{iA} g^{jB} \bigg] + \nonumber \\
  && \hspace*{8mm} 2 (\partial_i {\hat P}_A - \epsilon \partial_A {\hat P}_i)
    (\partial_j {\hat P}_B - \epsilon \partial_B {\hat P}_j)
    (g^{ij}g^{AB} - g^{iB} g^{jA}) + \nonumber \\
  && \hspace*{8mm} \epsilon {\hat F}_{AC} \bigg[
    4(\partial_i {\hat P}_B - \epsilon \partial_B {\hat P}_i) g^{iA} g^{BC}
    + \epsilon {\hat F}_{BD} g^{AB} g^{CD} \bigg] \bigg\}
    - {\frac14} (X^2-1)^2 \bbox{\Bigg)}. \label{lorder}
\end{eqnarray}
This now allows us to expand rigorously in powers of $\epsilon$. Note that so
far we have made no assumptions about any of the fields or their dependence on
the coordinates. We have simply rewritten the equations of motion, scaling with
respect to the physical dimensionful quantities in the problem, leaving the
equations of motion in terms of the dimensionless parameter $\epsilon$.

The procedure is as follows. We first consider the $\epsilon\to0$ limit,
with the metric and fundamental forms taking their geometrically 
defined values on the worldsheet. Since $\epsilon\to0$ is also the 
flat worldsheet limit, we expect that the fields will take the Nielsen--Olesen
form, which indeed turns out to be the case. In the next section
we will go to higher orders in $\epsilon$, deriving the corrections
to the geometry and fields, and hence the corresponding effective
action. However, for the moment we conclude this section by proving that the
Nielsen--Olesen solution is in fact the leading order behaviour for the
vortex fields around the worldsheet.

\begin{figure}
  \begin{center}
    \epsfig{file=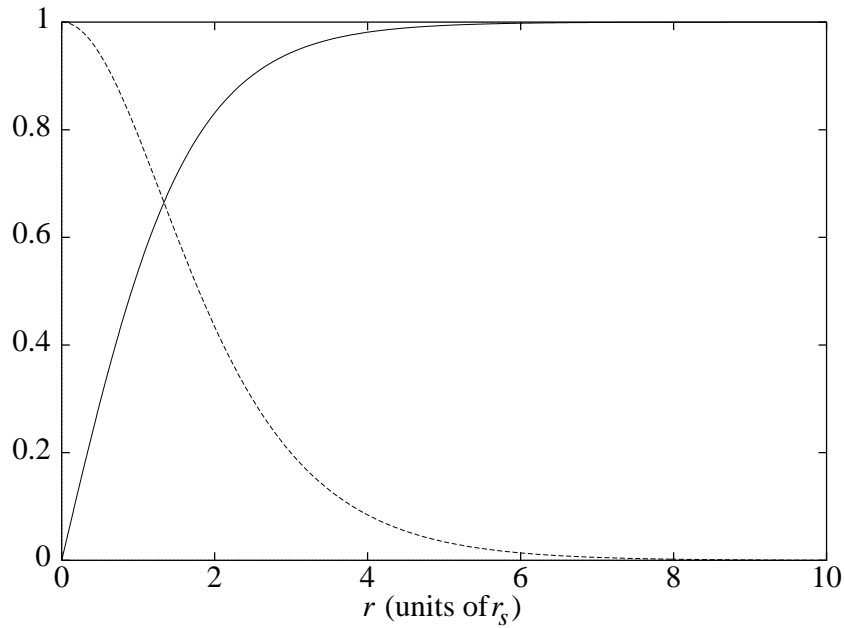,width=8.6cm,angle=270}
  \end{center}
  \caption{The Nielsen--Olesen solution for the critical case $\beta = 1.0$.
           This solution has been found using the relaxation methods (and
           routines) described in~\protect\cite{NRec}, by giving the
           conditions at $r = 0$ for $X$ (solid line) and $\hat P$ and at
           $r \to \infty$ for $X'$ and $\hat{P}'$.}
  \label{fig:NO}
\end{figure}

To zeroth order
\begin{equation}\label{liezero}
  {\cal L}_i h_{\mu\nu} = {\cal L}_i g_{\mu\nu} = {\cal L}_iK_{_j\mu\nu} = 0,
\end{equation}
hence all geometrical quantities take their worldsheet values,
$K_{_j\mu\nu}(s^A,x^i) = K_{_j\mu\nu}(s^A,0)$ etc. Hence
\begin{equation}\label{zerog}
  g_{_0\mu\nu} = \left( {\matrix{ \gamma_{AB} & 0 \cr 0 & -\delta_{ij} \cr}}
  \right)
\end{equation}
and thus the equations of motion for $X_{_0}$ and $P_{_0\mu}$ are
\begin{mathletters} \label{xpzero}
  \begin{eqnarray}
    -\partial_i\partial_i X + ({\hat P}_i^2 - {\hat P}_A^2) X
      + {\frac12}X(X^2-1) &= 0 \\
    - \partial_i{\hat F}_{ij} + \beta X^2 {\hat P}_j &= 0 \\
    - \partial_i\partial_i {\hat P}_A+ \beta X^2 {\hat P}_A &= 0.
  \end{eqnarray}
\end{mathletters}
These are solved by the Nielsen--Olesen solution, which is plotted in figure~%
\ref{fig:NO} against $r$ for $\beta = 1$:
\begin{equation}\label{solve}
  X_{_0} = X_{\rm NO}(r), \qquad {\hat P}_{_0j} = P_{\rm NO}(r)
  \partial_j \theta, \qquad {\hat P}_{_0A} = 0.
\end{equation}

Substitution of this solution back into~(\ref{actfour}) yields the familiar
Nambu action. It is the corrections to this action we are interested in.

\section{The corrections to the Nambu action} \label{sec:correct}

We now derive the leading order corrections to the Nambu action. 
In order to calculate quantities away from the worldsheet, such as
the metric, we must perform a Taylor expansion off the worldsheet
\begin{equation}\label{taylor}
  Q = Q|_0 + \xi^i{\cal L}_i Q|_0 + \frac12 \xi^i\xi^j{\cal L}_i {\cal L}_j
      Q|_0 + \frac16 \xi^i\xi^j\xi^k{\cal L}_i {\cal L}_j {\cal L}_k
      Q|_0 + \ldots
\end{equation}
However, in calculating the Lie derivatives of the metric and its
determinant some fortuitous cancellations occur. First note that
\begin{equation}\label{doublek}
  {\cal L}_k{\cal L}_j K_{_i\mu\nu} = \epsilon {\cal L}_k
  \left( K_{_i\sigma(\mu} K_{_j\nu)\rho} g^{\sigma\rho} \right) =0
\end{equation}
and also that 
\begin{eqnarray}
  {\cal L}_k{\cal L}_j{\cal L}_i \sqrt{-g} = \epsilon^3 \sqrt{-g} \; \big(
    && K_{_i}K_{_j}K_{_k} - K_{_i}K_{_j\nu}^\mu K_{_k\mu}^\nu -
    K_{j}K_{_i\nu}^\mu K_{_k\mu}^\nu - \nonumber \\
  && K_{_k}K_{_j\nu}^\mu K_{_i\mu}^\nu 
    + 2 K_{_i\sigma}^\mu K_{_j\mu}^\nu K_{_k\nu}^\sigma \big)
    = 0, \label{triplek}
\end{eqnarray}
which follows from a trace identity for the $2\times2$ matrix $K$~\cite{A}.
Hence to all orders, the metric and the volume jacobian are given by
\begin{mathletters}
\begin{eqnarray}
  g_{\mu\nu} &=& g_{\mu\nu}|_0 + 2 \epsilon x^i \left(
    K_{_i\mu\nu}|_0 - \varepsilon_{ij} \omega_{(\mu} n_{_j\nu)} |_0 \right)
    + \epsilon^2 x^i x^j \left( K_{_i\sigma(\mu}|_0 K_{_j\nu)}^\sigma |_0
    - \delta_{ij} \omega_\mu \omega_\nu \right) \\
  \sqrt{-g} &=& \sqrt{-g}|_0 \bigg[ 1 + \epsilon x^iK_{_i}|_0 + {\frac12}
    \epsilon^2 x^ix^j
    \left ( K_{_i}|_0 K_{_j}|_0 - K_{_iAB}|_0 K_{_j}^{AB}|_0
    \right ) \bigg].
\end{eqnarray}
\end{mathletters}
This gives all the information on the geometrical contributions to the
action. Now let us turn to the field theoretic contributions.

First note that since
\begin{equation}\label{triv}
  \delta S|_{g_{_0\mu\nu}} = \int d^2\sigma d^2x \sqrt{-\gamma} \:
    \left( {\delta S\over \delta X} \Big | _{g_{_0\mu\nu}} \hskip -2mm
    \delta X + {\delta S\over \delta P_\mu} \Big | _{g_{_0\mu\nu}} \hskip -2mm
    \delta P_\mu \right)
\end{equation}
vanishes by the equations of motion, first order field corrections contribute
at second order in the action, and second order corrections at fourth order.

To first order we may read off the equations for the 
corrections to the fields as
\begin{mathletters}\label{first}
  \begin{eqnarray}
    -\partial_i\partial_i X_{_1} - K_{_i}\partial_i X_{_0}
      + {\hat P}_{_0i}{\hat P}_{_0i} X_{_1} + 2 {\hat P}_{_0i}{\hat P}_{_1i}
      X_{_0} + {\frac12} X_{_1} (3 X_{_0}^2 -1) &=& 0 \label{first_a} \\
    - \partial_i{\hat F}_{_1ij} - K_{_i} {\hat F}_{_0ij}
      + \beta X_{_0}^2 {\hat P}_{_1j} + 2\beta X_{_0} X_{_1}
      {\hat P}_{_0j} &=& 0 \label{first_b} \\
    - \partial_i ( \omega_A x^j\varepsilon_{jk}{\hat F}_{ik}) -
      \partial_i \partial_i {\hat P}_{_1A} + \beta X_{_0}^2 {\hat P}_{_1A}
      + \beta X_{_0}^2 \omega_A x^j \varepsilon_{jk}{\hat P}_{_0k} &=& 0.
  \end{eqnarray}
\end{mathletters}
The first two equations do not apparently lend themselves to a straightforward
solution, however, we note that the equation of motion for  
the worldsheet obtained by varying the Nambu action is
\begin{equation}\label{nambueq}
  \Box X^\mu = n_{_i}^\mu K_i = 0,
% \sqcup\llap{$\sqcap$} X^\mu = n_{_i}^\mu K_i = 0,
\end{equation}
which would indicate that the `driving' terms in~(\ref{first_a}, \ref{first_b})
vanish and hence the appropriate solution is $X_{_1} = P_{_1 j} = 0$.
Indeed, \cite{GG} demonstrated that  unless the trace of the extrinsic
curvature vanished on the defect, the first order perturbation equations
have no solution that is regular and bounded. Thus in fact the appropriate
solutions to the first two equations of (\ref{first}) are $X_1=0$ and
$P_{_1j}=0$.

The last equation of~(\ref{first}) is solved by~\cite{A}
\begin{equation}\label{ptwist}
  {\hat P}_{_1A} = - \omega_A x^j\varepsilon_{jk} {\hat P}_{_0k};
\end{equation}
the presence of this term, as pointed out in~\cite{A}, 
guarantees the gauge invariance of the effective action to 
worldsheet SO(2) gauge transformations, for on
substituting in the form of the fields, this correction
to ${\hat P}_A$ exactly cancels the `twist' terms of Maeda and Turok,
and we arrive at the second order result
\begin{mathletters} \label{firstact}
\begin{eqnarray}
  S &=& - \mu \int d^2 \sigma \sqrt{-\gamma} \; \left( 1 
        - \epsilon^2 {\alpha_1\over\mu} K_{_iAB}^2 \right) \\
    &=& -\mu\int d^2 \sigma \sqrt{-\gamma} \; \left( 1
        - r_s^2 {\alpha_1\over\mu} {\cal K}_{_i\mu\nu}^2 \right),
\end{eqnarray}
\end{mathletters}
where
\begin{mathletters} \label{fcoeff}
 \begin{eqnarray}
   \mu &=& 2\pi\eta^2 \int_0^\infty rdr\hspace*{0.6mm} \bigg( X_{_0}^{\prime 2}
   + {X_{_0}^2 {\hat P}_{_0}^2 \over r^2} + {{\hat P}_{_0}^{\prime 2} \over r^2
   \beta} + {\frac14} (X_{_0}^2-1)^2 \bigg) \\
  \alpha_1 &=& {\pi\eta^2\over2} \int_0^\infty r^3dr \bigg( X_{_0}^{\prime 2} +
  {X_{_0}^2 {\hat P}_{_0}^2 \over r^2} + {{\hat P}_{_0}^{\prime 2}
  \over r^2 \beta} + {\frac14} (X_{_0}^2-1)^2 \bigg).
  \end{eqnarray}
\end{mathletters}

For the fourth order action we need the second order corrections to the fields, 
the volume factor already being exact to all orders.
From~(\ref{xorder}--\ref{paorder}) we may read off the equations for these as
\begin{mathletters} \label{xpsecond}
  \begin{eqnarray}
    -\partial_i\partial_i X_{_2} + x^jK_{_iAB}K_{_j}^{AB}\partial_i X_{_0}
      + {\hat P}_{_0i}{\hat P}_{_0i} X_{_2} + 2 {\hat P}_{_0i}{\hat P}_{_2i}
      X_{_0} + {\frac12} X_{_2} (3 X_{_0}^2 -1) &=& 0 \\
    - \partial_i{\hat F}_{_2ij} + x^kK_{_iAB}K_{_k}^{AB} {\hat F}_{_0ij}
      + \beta X_{_0}^2 {\hat P}_{_2j} + 2\beta X_{_0} X_{_2} {\hat P}_{_0j}
      &=& 0 \\
    - \partial_i ( \omega_A x^j\varepsilon_{jk}{\hat F}_{ik}) - \partial_i
      \partial_i {\hat P}_{_2A} &=& 0.
  \end{eqnarray}
\end{mathletters}
To simplify these, and to remove the explicit $K$-dependence we will
decompose in cylindrical harmonics by setting
\begin{mathletters} \label{secondh}
  \begin{eqnarray}
    X_{_2} &=& {\frac12} g K_{_iAB}^2 + {\bar g} x^{ij} K_{_iAB} K_{_j}^{AB} \\
    P_{_2\phi} &=& {\frac12}  q_\phi K_{_iAB}^2 + {\bar q_\phi}
      x^{ij} K_{_iAB} K_{_j}^{AB} \\
    P_{_2r} &=& q_r \varepsilon_{ki} K_{_kAB} K_{_j}^{AB}x^{ij},
  \end{eqnarray}
\end{mathletters}
where
\begin{equation}\label{xij}
  x^{ij} = {x^ix^j\over r^2} - {\frac12} \delta_{ij}.
\end{equation}
This gives two sets of coupled second order differential equations
\begin{mathletters} \label{zerohh}
  \begin{eqnarray}
    -g'' - {g'\over r} + {{\hat P}_{_0}^2 g \over r^2}
      + {2 {\hat P}_{_0} X_{_0}  q_\phi\over r^2} +
   {\frac12} g(3X_{_0}^2 -1) &=& -r X_{_0}' \\
    - q_\phi'' + { q_\phi'\over r} + \beta X_{_0}^2  q_\phi
      + 2 \beta X_{_0} {\hat P}_{_0}g &=& -r {\hat P}_{_0}'
  \end{eqnarray}
\end{mathletters}
and
\begin{mathletters} \label{secondeq}
  \begin{eqnarray}
    -{\bar g}'' - {{\bar g}'\over r} + {4{\bar g}\over r^2} 
      + {{\hat P}_{_0}^2 {\bar g} \over r^2}
      + {2 {\hat P}_{_0} X_{_0} {\bar q_\phi}\over r^2} +
    {\frac12} {\bar g}(3X_{_0}^2 -1) &=& -r X_{_0}' \\
    -{\bar q_\phi}'' + {{\bar q_\phi}'\over r} + 2 q_r' - {2 q_r\over r} 
      + \beta X_{_0}^2 {\bar q_\phi} +
    2 \beta X_{_0} {\hat P}_{_0}{\bar g} &=& -r {\hat P}_{_0}' \\
    4 q_r - 2{\bar q_\phi}' + \beta r^2 X_{_0}^2  q_r &=& - r^2 {\hat P}_{_0}'.
      \label{secondeq_c}
  \end{eqnarray}
\end{mathletters}
There are no additional corrections to ${\hat P}_A$ at this level.

In the critical case $\beta = 1$, the above equations reduce to two sets of
first order ODE's:
\begin{mathletters} \label{eq:crit}
  \begin{eqnarray}
    r g' &=& X_{_0} q_\phi + \hat{P}_{_0} g, \\
    \frac{q_\phi '}{r} &=& X_{_0} g + \hat{P}_{_0}
  \end{eqnarray}
\end{mathletters}
and
\begin{mathletters} \label{eq:crit_2}
  \begin{eqnarray}
    r \bar{g}' &=& X_{_0} \bar{q}_\phi + \hat{P}_{_0} \bar{g} +
      \frac{r \hat{P}_{_0}'}{X_{_0}}, \\
    2 \bar{g} &=& r q_r X_{_0} + \frac{r \hat{P}_{_0}'}{X_{_0}}, \\
    2 \bar{q}_\phi ' - 4q_r &=& 2 r \bar{g} X_{_0}.
  \end{eqnarray}
\end{mathletters}
Note that just as~(\ref{secondeq}) is really two coupled second order ODE's,~%
(\ref{eq:crit_2}) is really two coupled first  order ODE's.
The fact that~(\ref{zerohh}, \ref{secondeq}) reduce to such a first order
form for $\beta = 1$ is reassuring, since we do not expect to destroy
supersymmetry simply by bending the string!

Substituting~(\ref{secondh}) into the action, and after some algebra,
one arrives at Anderson's result: 
\begin{mathletters} \label{fourth}
  \begin{eqnarray}
    S &=& - \mu \int d^2 \sigma \sqrt{-\gamma} \bigg\{ 1
          - \epsilon^2 {\alpha_1\over\mu} K_{_iAB}^2 + \epsilon^4
          \bigg[ {\alpha_2\over\mu}(K_{_iAB}^2)^2 + {\alpha_3\over\mu}
          K_{_iAB}K_{_j}^{AB} K_{_iCD} K_{_j}^{CD}\bigg] \bigg\} \\
      &=& - \mu \int d^2 \sigma \sqrt{-\gamma} \bigg\{ 1
          - r_s^2 {\alpha_1\over\mu} {\cal K}_{_i\mu\nu}^2 \hspace{1mm} + r_s^4
          \bigg[ {\alpha_2\over\mu} ({\cal K}_{_i\mu\nu}^2)^2 +
          {\alpha_3\over\mu}{\cal K}_{_i\mu\nu}{\cal K}_{_j}^{\mu\nu} 
          {\cal K}_{_i\lambda\rho} {\cal K}_{_j}^{\lambda\rho}\bigg] \bigg\},
  \end{eqnarray}
\end{mathletters}
where
\begin{mathletters} \label{scoeff}
  \begin{eqnarray}
    \alpha_2 &=& {\pi\eta^2\over 4} \int_0^\infty dr \bigg[ 
      r^2X_{_0}'(2 g - {\bar g}) + {\hat{P}_{_0}'\over\beta} (2 q_\phi
      - {\bar q_\phi} - r q_r) + {4r\hat{P}_{_0}^2 \over\beta} \bigg] \\
    \alpha_3 &=& {\pi\eta^2\over 4} \int_0^\infty dr \bigg[ 
      2r^2X_{_0}'{\bar g} + {2\hat{P}_{_0}'\over\beta}({\bar q_\phi} + r q_r)
      - {4r\hat{P}_{_0}^2 \over\beta} \bigg].
  \end{eqnarray}
\end{mathletters}
If $\beta = 1$, it can be seen from~(\ref{eq:crit}, \ref{eq:crit_2}) that
$2 \alpha_2 + \alpha_3 = 0$.

\begin{figure}
  \begin{center}
    \epsfig{file=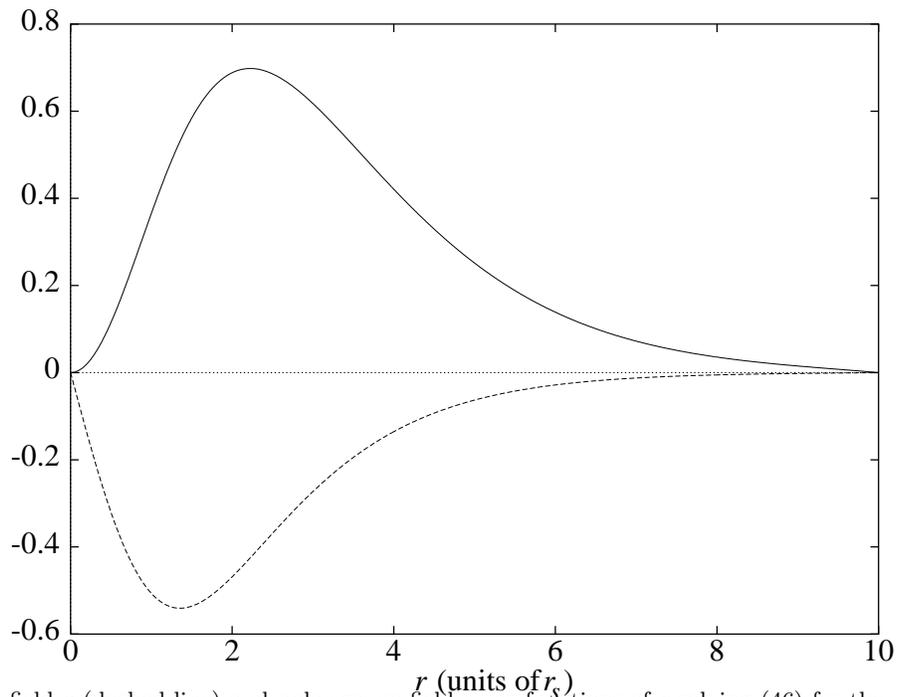,width=8.6cm,angle=270}
  \end{center}
  \caption{Higgs field $g$ (dashed line) and polar gauge field $q_\phi$
           as functions of $r$ solving (\protect\ref{zerohh}) for the value
           $\beta = 1.0$.}
  \label{fig:21}
\end{figure}

\begin{figure}
  \begin{center}
    \epsfig{file=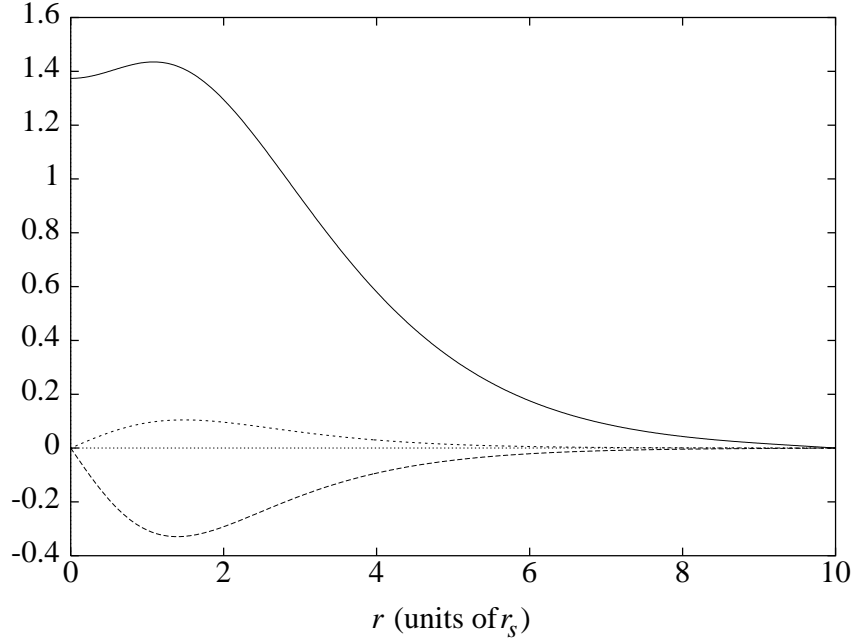,width=8.6cm,angle=270}
  \end{center}
  \caption{Higgs field $\bar g$ (dashed line), polar gauge field $\bar{q}_\phi$
           (solid line) and radial gauge field $q_r$ (in function of $r$)
           solving~(\protect\ref{secondeq}) for $\beta = 1.0$.}
  \label{fig:22}
\end{figure}

To determine $\alpha_2$ and $\alpha_3$, we need to solve
equations~(\ref{zerohh}) and~(\ref{secondeq}). This was done numerically by
relaxation methods, using NAG routine D02GAF~\cite{NAG}. Sample solutions
for both sets of equations for the critical case $\beta = 1$
are plotted in figures~\ref{fig:21} and~\ref{fig:22}. The first set of
equations can be readily implemented and was solved by requiring
that $g(0) = g(\infty) = q_\phi(0) = q_\phi(\infty) = 0$. Before solving the
second set of equations, we expressed $\bar{q}_r$ from~(\ref{secondeq_c}) as:
\begin{equation}
  q_r = \frac{2 \bar{q}_\phi' - r^2 \hat{P}_0'}%
             {4 + \beta r^2 X_0^2}.
\end{equation}
and eliminated it from the equations, which were then solved by asking that
$\bar{g}(0)=\bar{g}(\infty)=\bar{q}_\phi'(0)=\bar{q}_\phi(\infty)=0$.
The solutions obtained clearly behave at the origin as:
\begin{mathletters}
  \begin{eqnarray}
    g, \bar{g}, q_r &\sim& r, \\
    q_\phi &\sim& r^2, \\
    \bar{q}_\phi &\sim& \mathrm{const},
  \end{eqnarray}
\end{mathletters}
which can be checked to satisfy the asymptotic counterparts of~%
(\ref{zerohh}, \ref{secondeq}).

\begin{figure}
  \begin{center}
    \epsfig{file=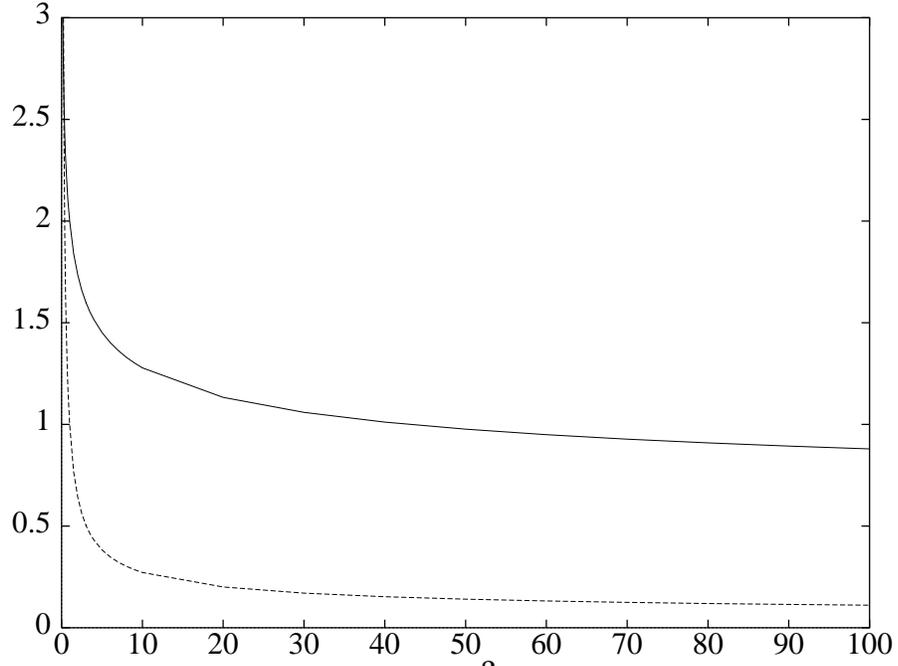,width=8.6cm,angle=270}
  \end{center}
  \caption{The parameters $\mu(\beta)/\pi\eta^2$ (solid line) and
           $\alpha_1(\beta)/\pi\eta^2$ appearing in the action to fourth
           order.}
  \label{fig:mualpha1}
\end{figure}

\begin{figure}
  \begin{center}
    \epsfig{file=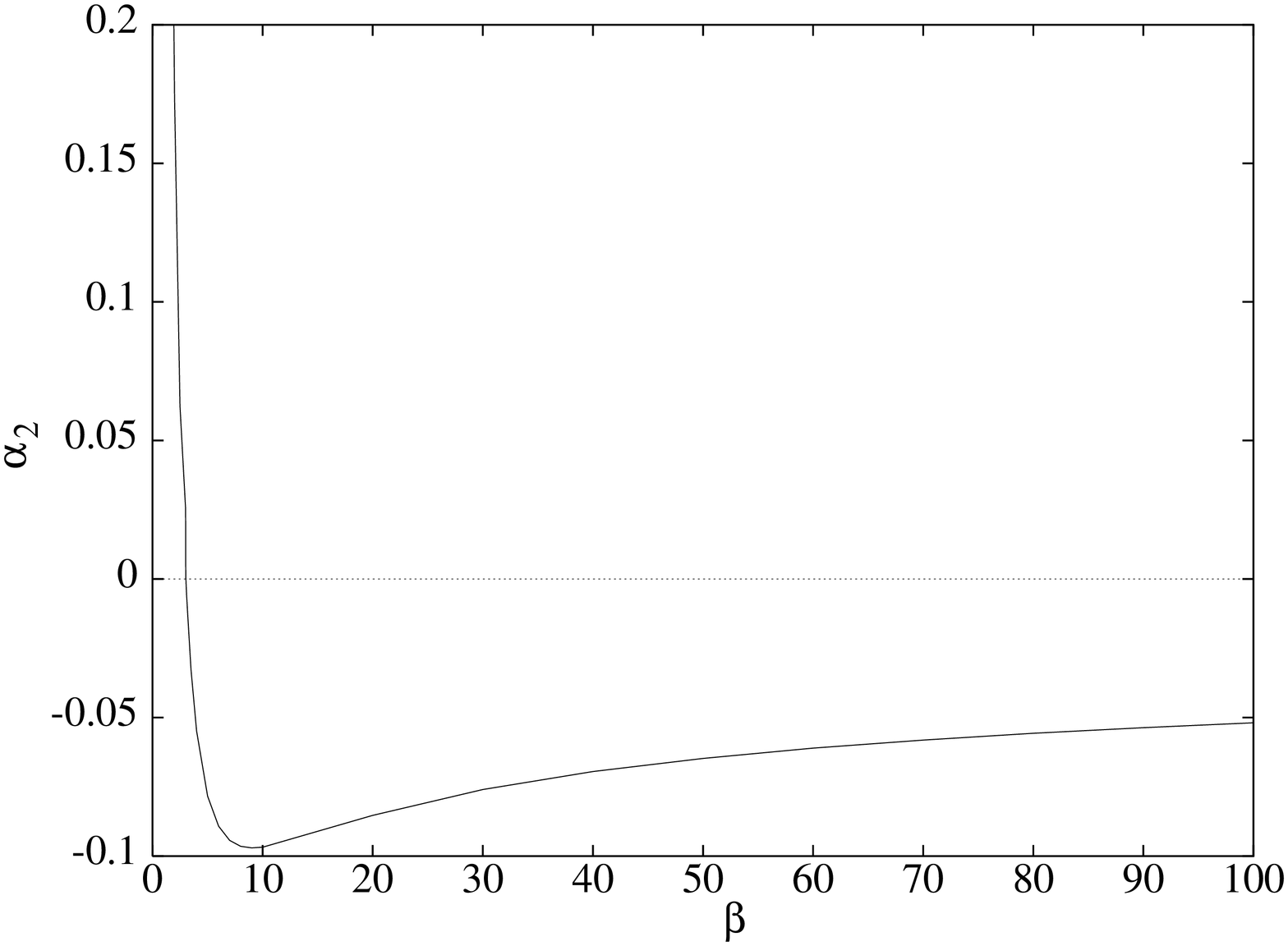,width=8.6cm,angle=270}
  \end{center}
  \caption{The parameter $\alpha_2(\beta)/\pi\eta^2$ appearing in the
           action to fourth order.}
  \label{fig:alpha2}
\end{figure}

\begin{figure}
  \begin{center}
    \epsfig{file=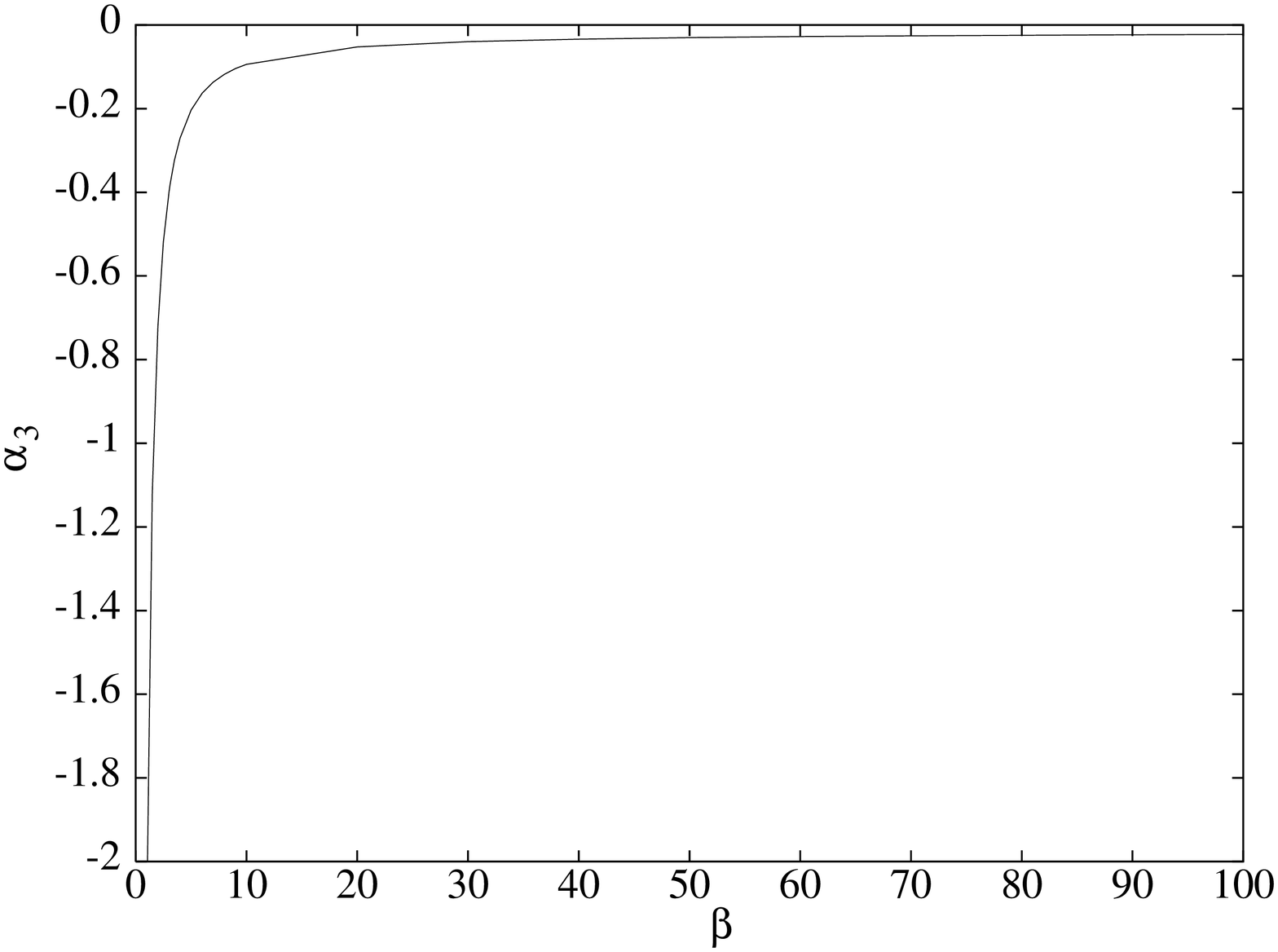,width=8.6cm,angle=270}
  \end{center}
  \caption{The parameter $\alpha_3(\beta)/\pi\eta^2$ appearing in the
           action to fourth order.}
  \label{fig:alpha3}
\end{figure}

We have computed the coefficients $\mu, \alpha_1, \alpha_2$ and $\alpha_3$ 
for a range of values of $\beta$; the results are shown in figures~%
\ref{fig:mualpha1}--\ref{fig:alpha3} and in table~\ref{table}. The key
points to note are that for the range of $\beta$ for which we computed
the coefficients ($0 < \beta \leq 100$) $\mu, \alpha_1 > 0$ and $\alpha_3,
\alpha_2 + \alpha_3 < 0$; however, $2 \alpha_2 + \alpha_3$ is positive
(negative) for $\beta < 1$ ($\beta > 1$) and $\alpha_2$ is positive for
$\beta < \beta_{\mbox{\footnotesize crit}} \simeq 3.03$ and negative for
$\beta > \beta_{\mbox{\footnotesize crit}}$.

\section{The equations of motion for the string} \label{sec:eom}

In order to derive the equations of motion for the string we must 
express the action~(\ref{fourth}) in terms of the worldsheet
coordinates, with respect to which we are varying the action.
To do this, we note that
\begin{equation}\label{malcone}
  K_{_iAB}  = n_{_i\mu},_A X^\mu,_B = -n_{_i\mu} X^\mu,_{AB}
  =-n_{_i\mu} X^\mu_{;AB}.
\end{equation}
Hence, defining
\begin{equation}\label{ndefn}
  N^{AB}_{CD} = X^{\mu;A}{}_{;C} X_{\mu;D}{}^{;B}
\end{equation}
we see that
\begin{mathletters} \label{trems}
  \begin{eqnarray}
    K_{_iAB}^2 &=& - X^{\mu;AB} X_{\mu;AB} = - N^{AB}_{BA} \\
    K_{_iAB}K_{_j}^{AB} K_{_iCD}K_{_j}^{CD} 
      &=& X^{\mu;AB}X_{\mu;CD} X^{\nu;CD} X_{\nu;AB} = N^{AC}_{BD} N^{DB}_{CA}.
  \end{eqnarray}
\end{mathletters}
Now, the connection on the worldsheet is given by
\begin{equation} \label{seven}
  \Gamma _{BC}^A = {\frac12} \gamma ^{AE}\left( \gamma _{BE},_C+\gamma
    _{EC},_B-\gamma _{BC},_E\right) = \gamma ^{AE}X^\mu,_E {X_\mu},_{BC}
\end{equation}
and the curvature (either directly or from the Gauss--Codazzi equations)
is
\begin{equation}\label{curv}
  R_{ABCD} = X^\mu_{;AC} X_{\mu;BD} - X^\mu_{;AD} X_{\mu;BC}.
\end{equation}
We may therefore, using the identity
\begin{equation}\label{twodr}
  R_{ABCD} = {\frac12} R \left(\gamma_{AC} \gamma_{BD} - \gamma_{AD} \gamma_{BC}
  \right)
\end{equation}
and the symmetries of $N$, infer the following useful relations:
\begin{mathletters} \label{userel}
  \begin{eqnarray}
    R^B_D &=& {\frac12} R \delta^B_D = - N^{AB}_{DA} \\
    R &=& - N^{AB}_{BA} \\
    N^{AC}_{BD} N^{DB}_{CA} &=& X^{\mu;AB} X_{\mu}^{\ ;CD} \left(
      R_{ACBD}+X^\nu_{;AD} X_{\nu;BC} \right)
      = {\frac12} R^2+N^{AD}_{BC} N^{CB}_{AD}.
  \end{eqnarray}
\end{mathletters}
We can now rewrite the action and its constituent variations as
\begin{equation}
  S = -\mu\int d^2\sigma\sqrt{-\gamma} \; \bigg\{ 1 - \epsilon^2
    {\alpha_1\over\mu} R + {\epsilon^4\over \mu} \label{news}
    \bigg[ (\alpha_2+{\frac12} \alpha_3) R^2 +
    \alpha_3 N^{AD}_{BC}N_{AD}^{CB}\bigg] \bigg\}
\end{equation}
and
\begin{mathletters} \label{vars}
  \begin{eqnarray}
    \delta \sqrt{-\gamma} &=& {\frac12} \sqrt{-\gamma} \: \gamma^{AB} \delta
        \gamma_{AB} = \sqrt{-\gamma} \: \gamma^{AB} X^\mu_{\ ,A}
        \delta X_{\mu,B} \\
    \delta R &=& - 2 X^{\mu;AB} \delta (X_{\mu;AB}) + 4N^{DA}_{AB} X^{\mu,B}
      \delta X_{\mu,D} = - 2  X^{\mu;AB}(\delta X_{\mu})_{;AB} -
      2 R X^{\mu,D} \delta X_{\mu,D} \\
    \delta \big(N^{AD}_{BC} N^{CB}_{AD}\big)
      &=& 4 (\delta X_\mu)_{;AB} X^{\mu;CD} 
      N^{AB}_{CD} - 4 X^\mu_{\ ,F} \delta X_{\mu,A} \gamma^{AF}
      N^{ED}_{BC} N^{CB}_{ED}
  \end{eqnarray}
\end{mathletters}
[using the fact that $\delta \Gamma ^C_{AB} X^\mu_{\ ,C}$ is parallel to
the worldsheet whereas $X^{\mu;AB}$ is perpendicular, and~(\ref{userel}).]

Noting that $\int \sqrt{-\gamma} R \propto \chi$, the Euler characteristic
which is a topological invariant, or from the above equations, we see that 
the $\alpha_1$ term will not contribute to the equations of motion.
The $R^2$ term, on the contrary, gives
\begin{equation} \label{rterm}
  3 \big( \gamma^{AB} X^\mu_{\ ,A} R^2 \big)_{;B}
    - 4 \left( R X^{\mu;AB} \right)_{;AB} =
  \left( X^{\mu,B} R^2 - 4 R_{,A} X^{\mu;AB} \right)_{;B} 
    = - 4 R_{;AB} X^{\mu;AB}
\end{equation}
where we used the Riemann identity
\begin{equation}\label{riemid}
  (D_A D_B - D_B D_A ) X^\mu{}_{,C} = R_{CDAB} X^{\mu,D},
\end{equation}
which implies
\begin{equation}\label{redriem}
  X^{\mu;AB}{}_{;B} = R^{AB} X^\mu_{\ ,B} = {\frac12} R X^{\mu,A}
\end{equation}
at each step.

The $N^2$ term gives
$$
  3 \left( X^{\mu,A} N^{ED}_{BC} N^{CB}_{ED} \right)_{;A}
  + 4 \left( N^{AB}_{CD} X^{\mu;CD} \right)_{;AB}\ ,
$$
but
\begin{eqnarray}
  N^{AB}_{CD;A} X^{\mu;CD} &=& X^{\nu;B}{}_{;DA} X_{\nu;C}{}^{;A}
    X^{\mu;CD} \nonumber \\
  &=& {\frac12} \left ( X^\nu_{\ ;DA} X_{\nu;C}^{\ \ \ ;A}
    \right ) _{;B} X^{\mu;CD} \nonumber \\
  &=& -{\frac14} \left ( R \gamma_{CD}\right )_{;B} X^{\mu;CD} = 0, \label{nid}
\end{eqnarray}
hence the full equations of motion for the system are:
\begin{equation} \label{fulleom}
  \frac\mu{\epsilon^4} \Box X^\mu = - (2\alpha_3 + 4 \alpha_2) R_{;AB}
    X^{\mu;AB} + 3 \alpha_3 X^{\mu,A} \left ( N^{ED}_{BC} N^{CB}_{ED}
    \right ) _{;A} + 4 \alpha_3 N^{AB}_{CD} X^{\mu;CD}{}_{;AB}.
\end{equation}
It is possible to express these equations in terms of the extrinsic curvatures.
Using~(\ref{malcone}) and
\begin{mathletters} 
\begin{eqnarray}
X^\mu_{;AB} &=& X^\mu_{,AB} - \Gamma^C_{AB} X^\mu_{,C} \label{covx} \\
X_{\mu,D}  X^\mu{}_{;AB} &=& 0, \label{scalx}
\end{eqnarray}
\end{mathletters}
it follows that
\begin{eqnarray}
  X^\mu_{;AB} &=& \eta^{\mu\nu} X_{\nu;AB}
               = \left(\gamma^{CD} X^\mu_{,C} X^\nu_{,D} -
                      \delta^{ij} n^\mu_{i} n^\nu_{j} \right) X_{\nu;AB}
                      \nonumber \\
              &=& - \delta^{ij} n^\mu_i n^\nu_j X_{\nu,AB}
               = \delta^{ij} n^\mu_i K_{_j AB}. \label{covx2}
\end{eqnarray}

Contracting the equation of motion with $n^\mu_{_i}$ and with $X^\mu_{,P}$
gives equations of motion normal and parallel to $\cal W$:
\begin{mathletters}
  \begin{eqnarray}
    - \frac\mu{\epsilon^4} K^A_{_iA} &=& (4 \alpha_2 + 2 \alpha_3) R_{;AB}
      K_{_i}^{AB} + 4\alpha_3 \delta^{jk} K_{_j}^{AC} K_{_k}^{BD} \; n^\mu_{_i}
      X_{\mu;CDAB} \label{eomperp} \\
    0 &=& 3 \alpha_3 \left(N^{CD}_{EF} N^{FE}_{CD} \right)_{,P} + 4 \alpha_3
      N^{CD}_{EF} \gamma^{AE} \gamma^{BF} \; X^\mu_{,P} X_{\mu;CDAB}
      \label{eompar}
  \end{eqnarray}
\end{mathletters}
Note that this latter equation~(\ref{eompar}) is an identity for the
unperturbed worldsheet. We have verified that it does indeed hold using
the lightcone gauge. Now,
\begin{eqnarray}
  n^\mu_{_i} X_{\mu;CDAB} = &-& K_{_i CD;AB} -
      \delta^{mn} \varepsilon_{im} \omega_B K_{_n CD;A}
      - \delta^{mn} \varepsilon_{im} \omega_A K_{_n CD;B} \nonumber \\
  &-& \delta^{mn} K_{_m CD} K^E_{_n A} K_{_i EB}
      - \delta^{mn} \varepsilon_{im} \omega_{A;B} K_{_n CD} + K_{_i CD}
      \omega_A \omega_B,
\end{eqnarray}
which implies that the equations of motion also read:
\begin{eqnarray}
  \frac\mu{\epsilon^4} K^A_{_i A} = - (4 \alpha_2 + 2 \alpha_3) R_{;AB}
     K_{_i}^{AB} + 4 \alpha_3 K_{_j}^{AC} K_{_j}^{BD} \big( &-& K_{_i CD;AB} -
    \varepsilon_{ik} \omega_B K_{_k CD;A} - \varepsilon_{ik}
    \omega_A K_{_k CD;B} \nonumber \\
  &-& K_{_k CD} K^E_{_k A} K_{_iEB} - \varepsilon_{ik} \omega_{A;B} K_{_k CD} +
    K_{_i CD} \omega_A \omega_B \big). \label{eomextr}
\end{eqnarray}

\section{Corrections to the motion of test trajectories} \label{sec:traj}

We now wish to derive the corrections to the motion of three
test trajectories:
the circular loop, the travelling wave, and the helical breather.
We have chosen these three examples because they provide
three different ways in which to observe the rigidity or otherwise of the 
string. 

The loop trajectory is given by
\begin{equation} \label{ltraj}
X^\mu(\tau,\sigma) = \left(\tau, \cos\tau\cos\sigma,
\cos\tau \sin\sigma, 0 \right),
\end{equation}
and collapses to a point after a time period $\Delta\tau = \pi/2 = L/4$,
where $L = \oint d\sigma = 2\pi$ is the length of the closed string.
The extrinsic curvature invariants become singular at this point, hence
rigidity would be indicated by a retardation of the collapse or a 
positive correction to the amplitude of the loop.

A travelling wave on the other hand is a variant of the flat worldsheet
where a deformation of arbitrary size and form is introduced, the
only constraint being that the deformation is a function of only
one of the lightcone coordinates, $\sigma_\pm = \sigma\pm\tau$:
\begin{equation} \label{ttraj}
X^\mu(\tau,\sigma) = \left(\tau, f(\tau-\sigma), g(\tau-\sigma),\sigma\right)
\end{equation}
This has been shown to be a solution of the full field theory~\cite{TW},
hence no correction should be found for this trajectory.

The helical breather is a time-dependent solution which may be written
as
\begin{equation} \label{htraj}
  X^\mu(\tau,\sigma) = \left(\tau, \sqrt{1-q^2}\cos\tau\cos\sigma,
  \sqrt{1-q^2}\cos\tau\sin\sigma, q\sigma\right)
\end{equation}
The limit $q\to1$ corresponds to the flat worldsheet and the limit
$q\to0$ to the collapsing circular loop. For intermediate $q$ the trajectory is
never singular, and the extrinsic curvature peaks at approximately
$q^{-2}\sqrt{1-q^2}$. Rigidity 
would be indicated by a preference for lower extrinsic
curvature and hence a negative correction to the amplitude of 
oscillation.

\subsection{Corrections to the motion of a collapsing loop}

As a first example of the effects of curvature terms on the motion of
freely-moving cosmic strings, we consider the collapse of a circular loop.
In Cartesian coordinates, the position of this string is given by
\begin{equation} \label{loop:position}
  X^\mu(\tau,\sigma) = \left(\tau, Z(\tau) \cos\sigma,
                       Z(\tau) \sin\sigma, 0 \right),
\end{equation}
and the normals to the worldsheet are
\begin{mathletters} \label{loop:normal}
  \begin{eqnarray}
    n_{_2}^\mu &=& \left(\dot{Z},\cos\sigma,\sin\sigma,0\right)/
                \sqrt{1 - \dot{Z}^2} \label{loop:normal1}\\
    n_{_3}^\mu &=& \left(0,0,0,1\right). \label{loop:normal2}
  \end{eqnarray}
\end{mathletters}

In this case, one obtains $\omega_A = 0$, and
\begin{mathletters}
  \begin{eqnarray}
  {\cal K}_{_2 AB} &=& \left( \begin{array}{c c}
                         \ddot{Z} & 0 \\
                            0     &  -Z
                       \end{array} \right) / \sqrt{1 - \dot{Z}^2}
                \label{loop:K2} \\
  {\cal K}_{_3 AB} &=& 0 \label{loop:K3} \\
  \gamma_{AB} &=& \left( \begin{array}{c c}
                         1 - \dot{Z}^2 & 0 \\
                               0       & -Z^2
                       \end{array} \right) \label{loop:metric} \\
  \Gamma^\tau_{AB} &=& \left( \begin{array}{c c}
                                \frac{-\dot{Z}\ddot{Z}}{1 - \dot{Z}^2} & 0 \\
                                0 & \frac{Z\dot{Z}}{1 - \dot{Z}^2}
                       \end{array} \right) \\
  \Gamma^\sigma_{AB} &=& \frac{\dot{Z}}{Z} \left( \begin{array}{c c}
                                                    0 & 1 \\ 1 & 0
                                                  \end{array} \right).
                         \label{loop:connect}
  \end{eqnarray}
\end{mathletters}

\noindent The equation of motion to zeroth order is
\begin{equation} \label{loop:zeroth}
  {\cal K}^A_{_2 A} = {Z \ddot{Z} + 1-\dot{Z}^2\over
    Z (1 - {\dot Z}^2)^{3/2}}= 0, \label{loop:eomzero}
\end{equation}
The general solution to this equation is $Z(\tau) = k\cos([\tau - \tau_0]/k)$.
Choosing $k = 1, \tau_0 = 0$ as initial conditions one obtains the canonical
form of the loop trajectory (\ref{ltraj}). Note that for this choice of 
loop length $\kappa = 1$ and we can use $K$ or ${\cal K}$ interchangeably.

We now wish to find the corrected solution to order $\epsilon^4 = r_s^4$.
To do this, we use equation (\ref{eomextr}) (technically its unrescaled
counterpart). Since $K_{_3AB}$ and $\omega^A$ vanish, the RHS of
(\ref{eomextr}) considerably simplifies to 
\begin{equation} \label{looprhs}
  4 ( \alpha_2 + \alpha_3) {\cal R}_{;AB} {\cal K}^{AB}
    = 32 (\alpha_2 + \alpha_3)\,\sec^8(\tau) \big(7\sec^2(\tau)-6\big).
\end{equation}
(Note that we have dropped the subscript `2' on the extrinsic curvature.)

\begin{figure}
  \begin{center}
    \epsfig{file=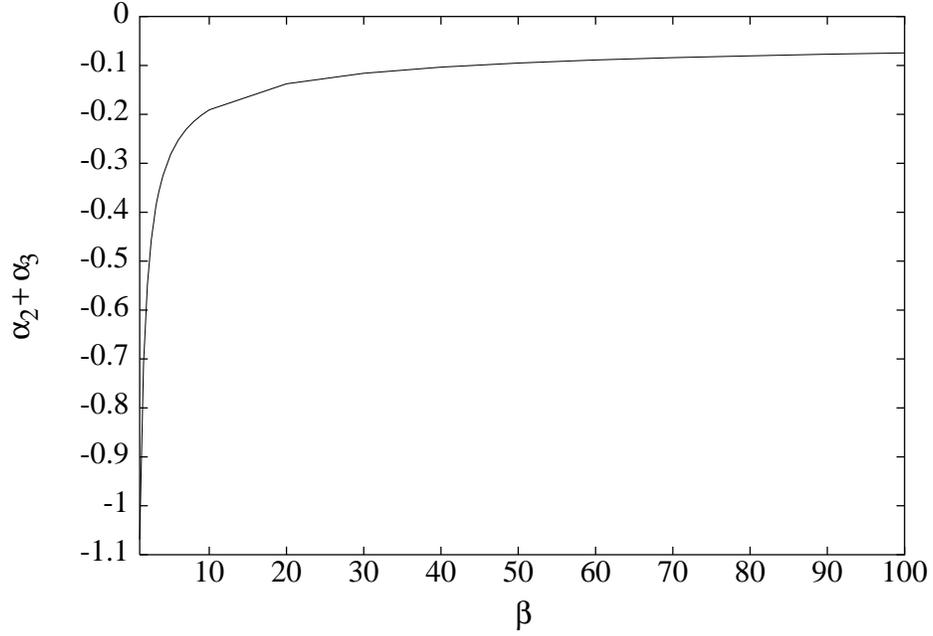,width=8.6cm,angle=270}
  \end{center}
  \caption{The `rigidity' parameter $(\alpha_2 + \alpha_3)(\beta)/\pi$
           appearing in the action to fourth order.}
  \label{fig:rigidity}
\end{figure}

\begin{figure}[htbp]
  \begin{center}
    \epsfig{file=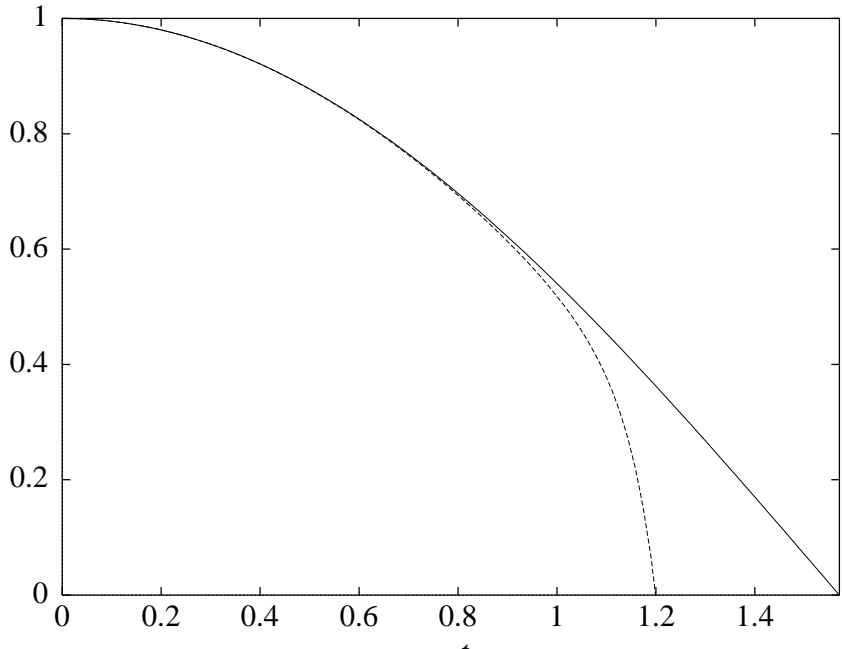,width=8.6cm,angle=270}
  \end{center}
  \caption{The collapse of a circular loop in the Nambu approximation
           (solid line) and at fourth order, for a (rather large) parameter
           $\epsilon = 1/10$.}
  \label{fig:collapse}
\end{figure}

The left-hand side is obtained by varying the trace of ${\cal K}_{_2}$ from
(\ref{loop:zeroth}), whereby we obtain
the equation for $\delta Z$ as
\begin{equation} \label{loop:fulleom}
  \delta \ddot{Z} + 2\tan(\tau) \delta \dot{Z} - \delta Z =
    32 \frac{r_s^4}{\mu} \left(\alpha_2 + \alpha_3 \right) \sec^5(\tau)
    \left(7\sec^2(\tau)-6\right).
\end{equation}
The solution to this, with initial conditions $\delta Z (0) =
{\dot {\delta Z}}(0)=0$ is
\begin{equation}
  \delta Z(\tau) = 32 \frac{\epsilon^4}{\mu} (\alpha_2 + \alpha_3)
    \bigg( \frac{7}{40}\sec^5(\tau) + \frac1{60}\sec^3(\tau)
    + \frac1{15}\sec(\tau) - \frac{31}{120}\cos(\tau) -
    \frac{\tau}{8} \sin(\tau) \bigg).
\end{equation}
Thus, the sign of $\alpha_2 + \alpha_3$ determines whether the string is rigid
or antirigid. This `loop rigidity parameter' is plotted against $\beta$ on
figure~\ref{fig:rigidity}, and its negativity means that the loop is antirigid
and tends to
collapse faster than in the Nambu approximation. This is illustrated on
figure~\ref{fig:collapse}, where we compare the zero-order solution
$Z(\tau) = \cos(\tau)$ with the corrected solution $Z(\tau) = \cos(\tau) +
\delta Z(\tau)$.

Note that the approximation breaks down when $|{\cal K}^A_B| = O(r_s^{-1})$,
i.e. when $\cos(\tau) = O(r_s)$. For example, in the illustration of
collapse in figure~\ref{fig:collapse}, $r_s=1/10$ is rather large; we would
hope that our approximation would be valid until the radii of
curvature of the worldsheet became close to 1/10, let us say twice the
radius of the string: 1/5. Inputting $|{\cal K}^A_B| = 1/5$ gives
$\tau \simeq 1.1$, which does indeed correspond to the point at which
the solutions start to significantly differ in figure~\ref{fig:collapse}.

\subsection{The motion of a travelling wave}

Now consider a travelling wave, whose position and normals are given by:
\begin{mathletters}
  \begin{eqnarray}
  X^\mu(\tau,\sigma) &=& \left(\tau,f(\tau-\sigma),g(\tau-\sigma),\sigma\right)
                         \label{wave:position} \\
  n_{_2}^\mu \equiv n^\mu &=& \left(0, g',-f',0\right) / \sqrt{f'{}^2+g'{}^2}
                         \label{wave:normal1} \\
  n_{_3}^\mu \equiv m^\mu &=& \left(f'{}^2 + g'{}^2, f', g',
                         f'{}^2 + g'{}^2 \right) / \sqrt{f'{}^2 + g'{}^2},
                         \label{wave:normal2}
  \end{eqnarray}
\end{mathletters}
where a prime denotes differentiation with respect to the argument, namely
$\tau - \sigma$.

Writing
\begin{mathletters}
  \begin{eqnarray}
    \lambda(\tau - \sigma) &=& f'{}^2 + g'{}^2 \\
    \zeta(\tau - \sigma)   &=& \left(f'' g' - f' g''\right) / \lambda \\
    \kappa_2(\tau - \sigma)&=& \left(f'' g' - f' g''\right) / \sqrt{\lambda}
                               \\
    \kappa_3(\tau - \sigma)&=& \left(f' f'' + g' g''\right) / \sqrt{\lambda},
  \end{eqnarray}
\end{mathletters}
one has:
\begin{mathletters}
  \begin{eqnarray}
  \omega_A &=& \zeta \left( \begin{array}{c} 1 \\ -1 \\ \end{array} \right)
               \label{wave:twist} \\
  K_{_i AB} &=& \kappa_i \left( \begin{array}{c c}
                                   1 & -1 \\ -1 & 1 \\
                                 \end{array} \right) \label{wave:curvatures} \\
  \gamma_{AB} &=& \left( \begin{array}{c c}
                           1 - \lambda & \lambda \\ \lambda & -1 - \lambda \\
                         \end{array} \right)
                         \label{wave:metric} \\
  \Gamma^A_{BC} &=& \frac{\lambda'}2 \left( \begin{array}{c} 1 \\ -1 \\
                                            \end{array} \right) \otimes
                                     \left( \begin{array}{c c}
                                              -1 & 1 \\ 1 & -1 \\
                                            \end{array} \right).
                    \label{wave:connection}
  \end{eqnarray}
\end{mathletters}
It is then straightforward to see that all terms on the right-hand side
of~(\ref{eomextr}) vanish separately, as do the traces of the extrinsic
curvatures: the travelling wave is an exact solution to (at least) fourth
order.

\subsection{Corrections to the motion of a helical string in breathing mode %
            \label{sec:breather}}

We now consider the string given by the following position functions:
\begin{mathletters}
  \begin{eqnarray}
  X^\mu &=& \left(\tau, Z(\tau) \cos(\sigma), Z(\tau)\sin(\sigma), q\sigma
            \right) \label{helical:position} \\
  n_{_2}^\mu &=& \left( 0, q\sin(\sigma), -q\cos(\sigma), Z
                             \right) / \sqrt{q^2 + Z^2}
                             \label{helical:normal1} \\
  n_{_3}^\mu &=& \left( \dot{Z}, \cos(\sigma), \sin(\sigma), 0
                             \right) / \sqrt{1 - \dot{Z}^2}.
                             \label{helical:normal2}
  \end{eqnarray}
\end{mathletters}
This string is helical with breathing $q$: the limits $q \to 0$ and
$q \to 1$ represent a collapsing loop and a straight string, respectively.
[Note how~(\ref{helical:position}--\ref{helical:normal2}) reduce to~%
(\ref{loop:position}--\ref{loop:normal}), and how all pertinent
quantities in the equations of motion are obtained from these expressions.]

With this choice of normals, the fundamental forms are:
\begin{mathletters}
  \begin{eqnarray}
    \gamma_{AB} &=& \left( \begin{array}{cc} 1 - \dot{Z}^2 & 0 \\ 0 &
                    - (q^2+Z^2) \\ \end{array} \right) \\
    K_{_2 AB}   &=& - \frac{q \dot{Z}}{\sqrt{q^2+Z^2}} \left( \begin{array}{cc}
                    0 & 1 \\ 1 & 0 \\ \end{array} \right) \\
    K_{_3 AB}   &=& \frac1{\sqrt{1 - \dot{Z}^2}} \left( \begin{array}{cc}
                    \ddot{Z} & 0 \\ 0 & - Z \\ \end{array} \right) \\
    \omega_A    &=& \frac{-q}{\sqrt{1 - \dot{Z}^2}\sqrt{q^2+Z^2}} \left(
                    \begin{array}{c} 0 \\ 1 \\ \end{array} \right).
  \end{eqnarray}
\end{mathletters}
Hence the equations of motion to zeroth order become
\begin{equation} \label{helical:eom0}
{\ddot{Z}\over(1 - {\dot Z}^2)^{3/2}} 
+ \frac{Z}{(q^2 + Z^2)(1 - {\dot Z}^2)^{1/2}} = 0.
\end{equation}
This equation admits for general solution
\begin{equation}
  Z(\tau) = \sqrt{k^2 - q^2} \cos \left( \frac{\tau - \tau_0}{k} \right),
\end{equation}
so that, choosing again the initial conditions $k = 1, \tau_0 = 0$ and calling
\begin{equation} \label{helical:Omega}
 \Omega(\tau) = \cos^2(\tau) + q^2 \sin^2(\tau),
\end{equation}
we have
\begin{mathletters}
  \begin{eqnarray}
  \gamma_{AB} &=& \Omega \left( \begin{array}{c c} 1 & 0 \\ 0 & -1
                               \end{array} \right) \label{helical:metric} \\
  K_{_2 AB} &=& q \sqrt{\frac{1-q^2}{\Omega}} \sin(\tau) \left(
                  \begin{array}{c c} 0 & 1 \\ 1 & 0 \\ \end{array} \right)
                  \label{helical:K2} \\
  K_{_3 AB} &=& - \sqrt{\frac{1-q^2}{\Omega}} \cos(\tau) \left(
                  \begin{array}{c c} 1 & 0 \\ 0 & 1 \\ \end{array} \right)
                  \label{helical:K3} \\
  \omega_A &=& -\frac{q}{\Omega} \left( \begin{array}{c} 0 \\ 1 \\ \end{array}
                               \right) \label{helical:twist}.
  \end{eqnarray}
\end{mathletters}

The right-hand side of the corrected equation of motion becomes then
$$
  - 32 \epsilon^4 \sqrt{1-q^2} \; \cos(\tau) \Omega^{-9/2}
    \left( \beta_1 - \beta_2 \Omega^{-1} + \beta_3 \Omega^{-2} -
    \beta_4 \Omega^{-3} \right),
$$
where
\begin{mathletters} \label{eq:betas}
  \begin{eqnarray}
    \beta_1 &=& 6 \left[ (\alpha_2 + \alpha_3) + \alpha_2 q^2 \right] \\
    \beta_2 &=& 7 (\alpha_2 + \alpha_3) + (38 \alpha_2 + 22 \alpha_3) q^2
                + (7  \alpha_2 - 3  \alpha_3) q^4 \\
    \beta_3 &=& 5 q^2 \left[ (7 \alpha_2 + 5 \alpha_3) +
                            (7 \alpha_2 + 2 \alpha_3) q^2 \right] \\
    \beta_4 &=& 15 q^4 (2 \alpha_2 + \alpha_3),
  \end{eqnarray}
\end{mathletters}
\noindent and the left-hand side is obtained as before by varying the trace of
$K_{_3}$:
\begin{equation}
  \delta \left( K^A_{_3 A} \right) = \Omega^{-3/2} \delta\ddot{Z} +
    2 \Omega^{-5/2} (1-q^2) \sin(\tau) \cos(\tau) \delta \dot{Z}
    + \Omega^{-5/2} \left[ q^2 - (1-q^2) \cos^2(\tau)\right]\delta Z.
    \label{helical:deltaK}
\end{equation}

Finally, the corrected equations of motion are:
\begin{eqnarray}
  \delta \ddot{Z} &+& 2\frac{1-q^2}{\Omega} \sin(\tau) \cos(\tau)\delta\dot{Z}
     + \frac{q^2 - (1-q^2) \cos^2(\tau)}{\Omega} \delta Z \nonumber \\
  &=& -32 \frac{\epsilon^4}{\mu} \sqrt{1-q^2} \cos(\tau)
    \Omega^{-3}\big( \beta_1 - \beta_2 \Omega^{-1} + \beta_3 \Omega^{-2}
    - \beta_4 \Omega^{-3} \big). \label{helical:fulleom}
\end{eqnarray}

Although it is possible to find an exact solution to this equation (see
appendix for details), it is
more instructive to consider the quasiflat limit $q \to 1$. In this case,
equation~(\ref{helical:fulleom}) becomes
\begin{equation}
  \delta \ddot{Z} + \delta Z = - 32 \frac{\epsilon^4 \Delta^3}{\mu}
    \big[  (\alpha_2 + \alpha_3) \cos(\tau) -
    (2\alpha_2+\alpha_3) \cos(\tau) \sin^2(\tau) \big],
\end{equation}
where $\Delta$ is defined by
\begin{equation} \label{eq:Delta}
  q^2 = 1 - \Delta^2.
\end{equation}
The solution $\delta Z(\tau)$ satisfying $\delta Z(0) = \delta \dot{Z}(0)
=0$ is then found to be
\begin{equation}
  \delta Z = - \frac{\epsilon^4 \Delta^3}{\mu} \bigg[ 4 \left( 2 \alpha_2 +
    3 \alpha_3 \right) \tau \sin(\tau) \\
    + ( 2 \alpha_2 + \alpha_3) \big(\cos\tau - \cos(3\tau) \big) \bigg]
\end{equation}
The corrected trajectory can then be written:
\begin{equation}
  Z + \delta Z = \Delta\left[ 1 - \frac{\epsilon^4 \Delta^2}{\mu} (2 \alpha_2 +
    \alpha_3) \right] \cos \left\{ \left[ 1 + \frac{4 \epsilon^4\Delta^2}%
    {\mu} ( 2 \alpha_2 + 3 \alpha_3 ) \right] \tau \right\}
    + \frac{\epsilon^4 \Delta^3}{\mu} (2 \alpha_2 + \alpha_3) \cos
    (3\tau)
  + \mbox{O}(\epsilon^8).
\end{equation}
The effect of the correction is threefold: First, it alters the frequency of 
the motion, $\tau \to \left[ 1 + 4 \epsilon^4 \Delta^2 (2\alpha_2 + 3\alpha_3)
/ \mu \right] \tau$; since $(\alpha_2 + \alpha_3), \alpha_3
< 0$ this has the effect of reducing the frequency---a tendency we would be
tempted to call rigid. Secondly, the amplitude of the oscillation is altered
by a factor $1 - \epsilon^4 \Delta^2 (2 \alpha_2 + \alpha_3) / \mu$.
This could be either a amplification or reduction,
depending on whether $\beta > 1$ or $\beta < 1$. Finally, a higher frequency
oscillation is superposed on the motion for $\beta \neq 1$. If, for 
simplicity, we take $\beta = 1$, so $2\alpha_2 + \alpha_3 = 0$, we see that
\begin{equation}
  Z + \delta Z = \Delta \cos \left[ \left( 1 + \frac{8\epsilon^4 \Delta^2}{\mu}
    \alpha_3 \right) \tau \right]
\end{equation}
i.e. the only effect of the correction is to reduce the frequency of
oscillation of the breather, which would seem to be unambiguously rigid.
 
However, we now observe a curious property: Suppose instead we consider 
initialising the correction at the instant of maximal velocity:
$\delta Z(-\pi/2) = \delta \dot{Z} (-\pi/2) = 0$, we find
\begin{equation}
  Z + \delta Z = \Delta \left[ 1 - \frac{\epsilon^4 \Delta^2}{\mu}
    ( 2 \alpha_2 + 9 \alpha_3) \right] \sin \left\{ \left[ 1 + \frac{4
    \epsilon^4 \Delta^2}{\mu} (2 \alpha_2 + 3 \alpha_3) \right] \tau'
    \right\} - \frac{\epsilon^4 \Delta^3}{\mu} (2 \alpha_2 + \alpha_3)
    \sin(3 \tau') + \mbox{O}(\epsilon^8)
\end{equation}
(where $\tau' = \tau + \pi/2$). Now note that while the frequency of
oscillation is decreased by the same amount, and the higher frequency term is
still the same, the amplitude is now uniformly increased for all $\beta$.
If, as before, we take $\beta = 1$, we now find
\begin{equation}
  Z + \delta Z = \Delta \left( 1 - \frac{8\epsilon^4 \Delta^2}{\mu} \alpha_3
    \right) \sin \left[ \left( 1 + \frac{8\epsilon^4 \Delta^2}{\mu} \alpha_3
    \right) \tau' \right],
\end{equation}
in other words an {\it increase\/} in the amplitude of oscillation accompanies
a similar decrease in its frequency. One cannot immediately see from this
solution
whether or not the behaviour is rigid, however, an analysis of the Ricci
curvature near $\tau' = 0$ shows that it is in fact increased---a behaviour
consistent with antirigidity! A calculation of the dependence of the curvature
on general $q$ is presented in the appendix. The algebra is more complicated
but the results are the same: all helical breathers display both rigid and
antirigid characteristics.
 
The results of our helical breather calculations therefore appear rather
ambiguous. If we wish to identify rigidity by the
behaviour of a corrected trajectory---whether it increases or decreases
curvature---we are forced to calculate the effect on the curvature (see
appendix) and then the results appear to depend on the initial conditions.
What this shows is that the `decrease in curvature' criterion for rigidity
is too na\"{\i}ve to be reliably applied in all situations.

\section{The question of rigidity} \label{sec:rigid}

\begin{figure}
  \begin{center}
    \epsfig{figure=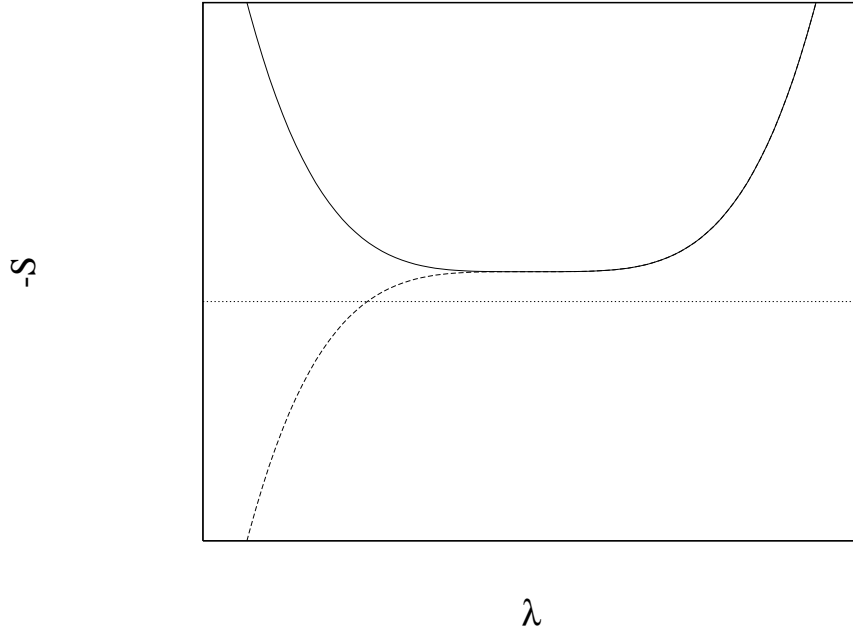,width=8.6cm,angle=270}
    \caption{Schematic graph of $-S(\lambda)$, the variation of the action
             as the worldsheet is rescaled. The solid line would correspond
             to a rigid string, and the dashed line to an antirigid string.
             \label{fig:rescale}}
  \end{center}
\end{figure}

We now want to determine whether a string can be labelled `rigid' or
`antirigid'. To do this, we use an argument based in that of
Polyakov~\cite{Po}, which consists in determining how the action varies under
a rescaling of the spacetime coordinates $X^\mu \to \lambda X^\mu$.
Such transformations alter the scale of crinkles of the worldsheet and
magnify or reduce small-scale structure, hence rigidity
would be indicated by a extremum of the energy or the action with respect to
the rescaling parameter $\lambda$, as illustrated in figure~\ref{fig:rescale}.

Out starting point is the action~(\ref{fourth}), which can be written
\begin{equation}
  -S = \mu \int d^2 \sigma \sqrt{-\gamma}
      - \epsilon^2 \alpha_1 \int d^2 \sigma \sqrt{-\gamma} \: M_{ii}
      + \epsilon^4 \alpha_2 \int  d^2 \sigma \sqrt{-\gamma} \: M^2_{ii}
      + \epsilon^4 \alpha_3 \int  d^2 \sigma \sqrt{-\gamma} \: M_{ij} M_{ij},
\end{equation}
where the matrix $M_{ij}$ is defined as
\begin{equation}
  M_{ij} = K_{_i AB} K_{_j}^{AB}.
\end{equation}

As $M_{ij} M_{ij} = M_{ii}^2 - 2 \, \det(M)$, this can be expressed as
\begin{equation}
  -S = \mu A -  \epsilon^2 \alpha_1 \chi +  \epsilon^4
  \big [ (\alpha_2 +\alpha_3) I_1 - 2 \alpha_3 I_2 \big ],
\end{equation}
with $A$ the area of the worldsheet for the range of $\{\tau, \sigma\}$
being integrated over and $\chi$ proportional to the Euler character of
$\cal W$. Also,
\begin{mathletters}
  \begin{eqnarray}
    I_1 &=& \int d^2 \sigma \sqrt{-\gamma} \: M^2_{ii}, \\
    I_2 &=& \int d^2 \sigma \sqrt{-\gamma} \: \det(M).
  \end{eqnarray}
\end{mathletters}

We now rescale $X^\mu \to \lambda X^\mu$, so that for $\lambda > 1$ the
worldsheet is expanded and for $\lambda < 1$ it is shrunk. Then,
\begin{mathletters}
  \begin{eqnarray}
    \gamma_{AB} &\to& \lambda^2 \gamma_{AB}, \\
    K_{_i AB} &\to& \lambda K_{_i AB},
  \end{eqnarray}
\end{mathletters}
and thus $I_i \to \lambda^{-2} I_i$, $i = 1, 2$. The shape of the curve
$S(\lambda)$ now depends explicitly on the integrals $I_i$, since $-S$ is
rescaled as
\begin{equation}
  -S \to \lambda^2 \mu A + \epsilon^2 \alpha_1 \chi
  + \lambda^{-2}  \epsilon^4 \big [
  (\alpha_2 +\alpha_3) I_1 - 2 \alpha_3 I_2 \big ].
\end{equation}

We know that $\alpha_3, \alpha_2 + \alpha_3 < 0$, and clearly $I_1 > 0$, so in
order to determine the shape of $S(\lambda)$ we only need to determine
the sign of $\det(M)$. For these purposes, we can work in the
conformal gauge, $\gamma_{AB} = \eta_{AB}$, and we find
\begin{equation}
  \det(M) =   \big( K_{_200} K_{_311} - K_{_300} K_{_211} \big)^2
          - 2 \big( K_{_211} K_{_310} - K_{_311} K_{_210} \big)^2 \nonumber \\
          - 2 \big( K_{_200} K_{_310} - K_{_300} K_{_210} \big)^2.
\end{equation}
If we impose the Nambu equations of motion, $K^A_{_i A} = 0$, this determinant is
strictly negative. Hence, since $\mu A$ is positive, we see that $S(\lambda)$
is unbounded, because the the coefficients multiplying $\lambda^2$ and
$\lambda^{-2}$ have opposite signs.

We must therefore conclude that the string is antirigid. This does not mean
that \emph{all} the trajectories of the string exhibit antirigidity, but
rather that it is impossible for all trajectories to be rigid.

Let us illustrate this by considering the trajectories of the previous section.
For the collapsing loop, $\det(M) = 0$, and as we noticed, $\alpha_2 +
\alpha_3$ determines alone the shape of $S(\lambda)$. This simplification
came from the fact that the loop is flat, and therefore has only one
nonvanishing extrinsic curvature.

For the travelling wave, $M_{ij} \equiv 0$, which is consistent with our
observation of no corrections to this trajectory.

In the case of the helical breather, both $M_{ii}$ and $\det(M)$ are non-zero,
so we do not expect results to depend on $\alpha_2 + \alpha_3$. Even though
the action is unstable to the scaling of the worldsheet, what is happening
with the helical breather is that the correction does not always have a 
non-zero projection on this unstable mode. However, we would expect a
general correction of the breather to exhibit an instability.

To sum up: We have reviewed the derivation of the effective action for a
U(1) local cosmic string to fourth order in the ratio of string width to
worldsheet curvature. We presented numerical results calculating the
coefficients  of these fourth order terms. We then derived the equations of
motion for the string to fourth order, and calculated corrections to a sample
of well-known trajectories. We have given a general argument for antirigidity
of the cosmic string to fourth order, however by reference to our examples have
shown that not all trajectories need behave in an `antirigid' fashion---%
rigidity it appears is rather like a theorem, it may work in special cases,
but one needs only find a single counterexample to disprove it.

\section*{Acknowledgements}

The authors would like to thank Lyndon Woodward for useful discussions and
Bernard Piette for help with numerical aspects of this work. RG was supported 
by the Royal Society and FB was partially supported by a grant from the Board
of the Swiss Federal Institutes of Technology.

\appendix
 
\section*{The Helical Breather}
 
Recall from section~\ref{sec:breather} that the helical breather solution
to the Nambu action is
\begin{equation}
  X^\mu = \big( \tau, \Delta \cos(\tau)\cos(\sigma), \Delta \cos(\tau)
          \sin(\sigma), q \sigma \big)
\end{equation}
[where $\Delta$ is defined in~(\ref{eq:Delta})] and that the equation of motion
for $\delta Z (\tau)$ is~(\ref{helical:fulleom}):
\begin{eqnarray}
  \delta \ddot{Z} &+& 2\frac{\Delta^2}{\Omega} \sin(\tau) \cos(\tau)\delta
     \dot{Z} + \frac{q^2 - \Delta^2 \cos^2(\tau)}{\Omega} \delta Z \nonumber \\
  &=& -32 \frac{\epsilon^4}{\mu} \Delta \cos(\tau)
    \Omega^{-3}\big( \beta_1 - \beta_2 \Omega^{-1} + \beta_3 \Omega^{-2}
    - \beta_4 \Omega^{-3} \big).
\end{eqnarray}

This can be solved using the method of variation of parameters, giving (after
a long and tedious calculation):
\begin{eqnarray}
  \delta Z(\tau) &=& \frac{16}{\mu \Delta} \left[ \left(
    \frac{\beta_1}3 - \frac{\beta_2}4 + \frac{\beta_3}5 - \frac{\beta_4}6
    \right) \left( \cos(\tau)+ \Delta^2 \tau \sin(\tau) \right) -
    \left(
     \frac{\beta_1}{3\Omega^3} - \frac{\beta_2}{4\Omega^4} +
    \frac{\beta_3}{5\Omega^5} - \frac{\beta_4}{6\Omega^6} \right)
    \cos(\tau) \right] \nonumber \\
  && + \lambda_0 \sin(\tau)\tan^{-1} \big( q \tan(\tau) \big)
     + \sin^2(\tau)\cos(\tau) \sum_{n=1}^{6} \lambda_n \Omega^{-n}.
\end{eqnarray}
Here, $\Omega(\tau) = \cos^2(\tau) + q^2 \sin^2(\tau)$, the $\beta$'s were
defined in~(\ref{eq:betas}) and the $\lambda$'s are:
\begin{mathletters}
\begin{eqnarray}
  \lambda_0 &=& \frac{\Delta}{240 q \mu} \bigg[
    640 \big( 3+1/q^2 \big) \beta_1 - 360 \big( 5+2/q^2+1/q^4 \big) \beta_2 +
    48 \big( 35 + 15/q^2 + 9/q^4 + 5/q^6 \big) \beta_3 \nonumber \\
  && \hspace*{38mm} -25 \big( 63+28/q^2+18/q^4+12/q^6+7/q^8 \big) \beta_4
    \bigg], \\
  \lambda_1 &=& \frac{\Delta}{720 \mu} \bigg[ 1920 \big(3-1/q^2\big) \beta_1 -
    360 \big( 15-4/q^2-3/q^4 \big) \beta_2 + 48 \big( 105-25/q^2-17/q^4-15/q^6
    \big) \beta_3 \nonumber \\
  && \hspace*{38mm} - 5 \big( 945-210/q^2-136/q^4-110/q^6-105/q^8 \big) \beta_4
    \bigg], \\
  \lambda_2 &=& \frac{\Delta}{360 \mu} \bigg[ 1920 \beta_1 - 360 \big( 5-1/q^2
    \big) \beta_2 + 48 \big( 35-6/q^2-5/q^4 \big) \beta_3 -
    5 \big( 315-49/q^2-39/q^4-35/q^6 \big) \beta_4 \bigg], \\
  \lambda_3 &=& \hspace*{0.9mm} \frac{\Delta}{90 \mu} \hspace*{0.9mm}
    \bigg[ 480 \beta_1 -360 \beta_2 +48 \big(
    7-1/q^2 \big) \beta_3 - 5 \big( 63-8/q^2 - 7/q^4 \big) \beta_4 \bigg], \\
  \lambda_4 &=& \hspace*{0.9mm} \frac{\Delta}{15 \mu} \hspace*{0.9mm}
    \bigg[ -60 \beta_2 + 48 \beta_3 -
    5 \big( 9 - 1/q^2 \big) \beta_4 \bigg], \\
  \lambda_5 &=& \hspace*{0.9mm} \frac{8 \Delta}{15 \mu} \hspace*{0.9mm}
    \bigg[ 6 \beta_3 - 5 \beta_4 \bigg], \\
  \lambda_6 &=& - \frac{8 \Delta}{3 \mu} \beta_4.
\end{eqnarray}
\end{mathletters}
 
To use this general solution to investigate the (anti)rigid nature
of these helicoidal trajectories, we need to observe how the Ricci curvature
\begin{equation}
  R = K_{_i AB} K_{_i}^{AB} - K_{_i} K_{_i}
    = - \frac{2q^2 \dot{Z}^2}{(q^2+Z^2)^2(1-\dot{Z}^2)} -
        \frac{2 Z \ddot{Z}}{(1-\dot{Z}^2)^2 (q^2+Z^2)}
\end{equation}
depends on the correction. For simplicity, we take $\beta = 1$, and note that
for the background solution
\begin{equation} \label{eq:appcurv1}
  R = - \frac{2 \Delta^2}{\Omega^3} \left( q^2 \sin^2(\tau) - \cos^2(\tau)
      \right).
\end{equation}

\begin{figure}
  \begin{center}
    \epsfig{figure=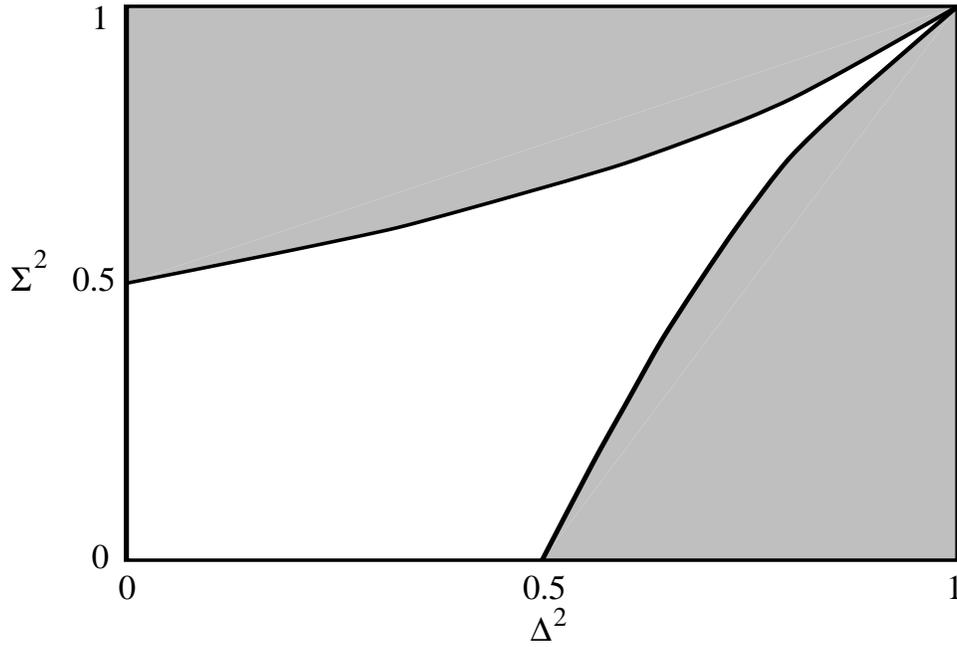,width=8.6cm,angle=270}
    \caption{Schema showing the regions of antirigidity (shaded) and rigidity
           for the helical breather at $\beta = 1$. Here $\Delta^2 =
           1 - q^2$ and $\Sigma = \sin(\tau_0)$. \label{fig:curv}}
  \end{center}
\end{figure}

Now, suppose we wish to investigate the behaviour of the curvature near a
general initial point $\tau_0$, where $\delta Z(\tau_0) = \delta \dot{Z}
(\tau_0) = 0$. Then near $\tau_0$
\begin{equation}
  \delta R \simeq - \frac{2 \Delta \cos(\tau_0)}{\Omega^3(\tau_0)} \delta
             \ddot{Z} (\tau_0)
           =  \frac{64 \epsilon^4 \Delta^2}{\mu} \frac{\cos^2(\tau_0)}%
               {\Omega^8(\tau_0)} \left( \beta_1 \Omega^2 - \beta_2 \Omega 
               + \beta_3 \right), \label{eq:appcurv2}
\end{equation}
where we have used~(\ref{helical:fulleom}) to evaluate $\delta \ddot{Z}
(\tau_0)$, and noted that $\beta_4 = 0$ for $\beta = 1$. Now, the combination 
$R \, \delta R$ will be negative if if the magnitude of the curvature is
decreased, which corresponds to an intuitive notion of rigidity.
From~(\ref{eq:appcurv1}) and~(\ref{eq:appcurv2}) we see that this requires:
\begin{equation}
  \left[ \left( 2 - \Delta^2 \right) \Sigma^2 - 1 \right] \\
  \left( - \Delta^2 + 2 \Delta^4 - 8 \Delta^4 \Sigma^2 +
    13 \Delta^6 \Sigma^2 - 6 \Delta^6 \Sigma^4 \right) > 0
\end{equation}
where $\Sigma^2 = \sin^2(\tau_0)$. Figure~\ref{fig:curv} shows the sign of
$R \, \delta R$ as a function of the
two parameters $\Sigma$ and $\Delta$. The shaded zones indicate the regions
where the string is antirigid. We see that---with the exception of the loop
case ($\Delta^2 = 1$)---the string admits both rigid and antirigid behaviour
for each value of $\Delta$.

\begin{figure}
  \begin{center}
    \epsfig{file=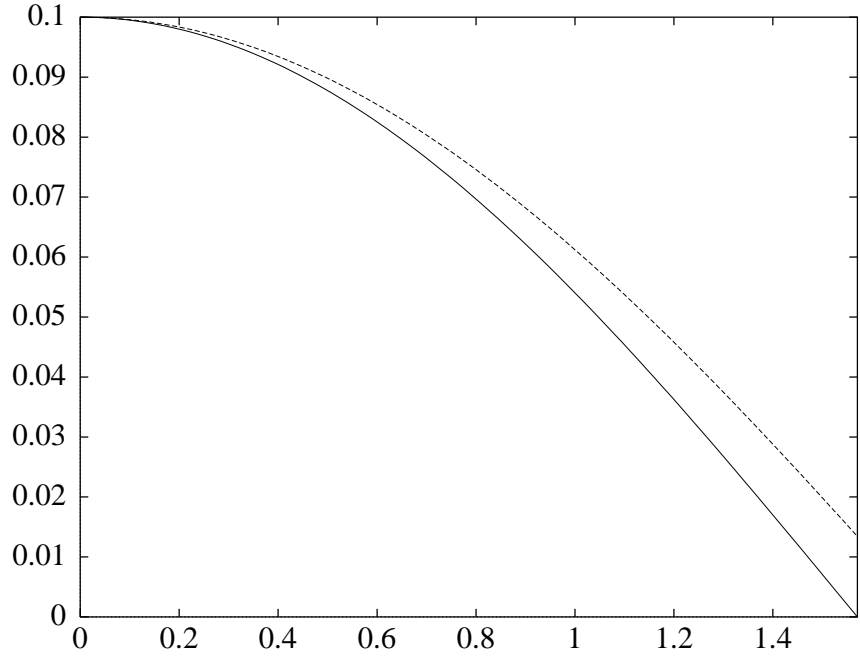,width=8.6cm,angle=270}
  \end{center}
  \caption{The corrected evolution of the quasiflat limit of a helical breather
           for $\beta = 1$
           and $\epsilon = \Delta = 1/10$. The correction added to
           the Nambu solution $Z(\tau)$ (solid line) is in fact $10^4 \,
           \delta Z$ in this figure (dashed line).}
  \label{fig:helical}
\end{figure}
 
\begin{figure}
  \begin{center}
    \epsfig{file=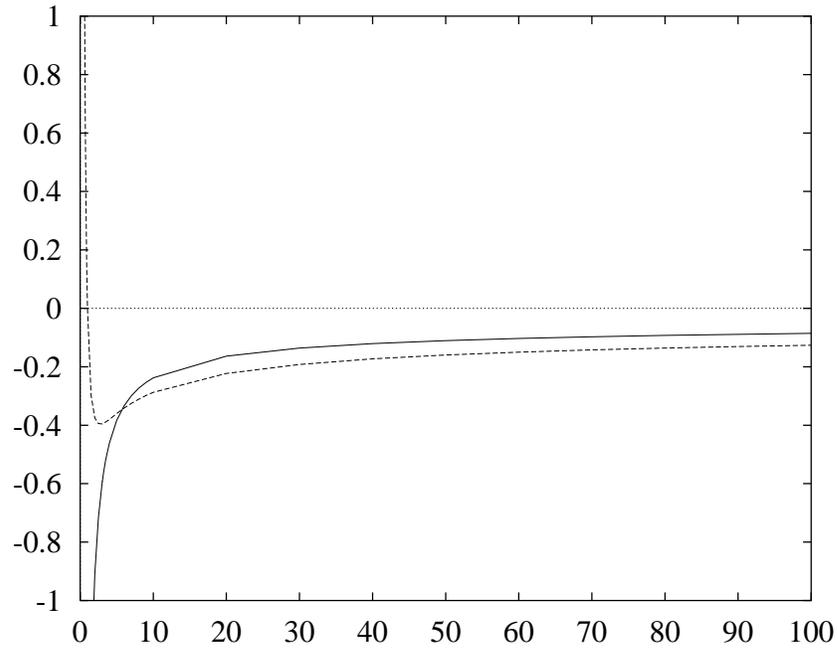,width=8.6cm,angle=270}
  \end{center}
  \caption{The coefficients $\alpha_2 + \frac32 \alpha_3$ (solid line) and
           $2 \alpha_2 + \alpha_3$ appearing in the solution $\delta Z$.
           Note that this last combination also appears in the action, and
           that it vanishes for the critical coupling $\beta = 1$.}
  \label{fig:helical2}
\end{figure}

\begin{table}
  \caption{The numerical coefficients appearing in the action to fourth order
           for some values of the Bogomol'nyi parameter $\beta$.}
  \label{table}
  \begin{tabular}{ddddd}
    $\beta$ & $\mu/\pi\eta^2$ & $\alpha_1/\pi\eta^2$ &
               $\alpha_2/\pi$ & $\alpha_3/\pi$ \\
    \tableline
      0.1 & 3.272 & 6.025 & 77.761 & -104.985 \\
      0.2 & 2.813 & 3.347 & 23.565 &  -32.835 \\
      0.5 & 2.314 & 1.634 &  4.308 &   -6.767 \\
      1.0\tablenotemark[1]
          & 2.000 & 0.999 &  1.069 &   -2.138 \\
      2.0 & 1.736 & 0.642 &  0.174 &   -0.722 \\
     10.0 & 1.278 & 0.272 & -0.097 &   -0.094 \\
     50.0 & 0.977 & 0.141 & -0.065 &   -0.030 \\
    100.0 & 0.880 & 0.111 & -0.052 &   -0.023 \\
  \end{tabular}
  \tablenotetext[1]{For $\beta = 1$, it can be analytically deduced from the
                    equations of motion that $\mu/\pi\eta^2 = 2
                    \alpha_1/\pi\eta^2 = 2$
                    and that $2 \alpha_2 + \alpha_3 = 0$.}
\end{table}


\begin{references}
  \bibitem[*]{mand}             E-mail: manderso@scorpion.cowan.edu.au
  \bibitem[\dagger]{fbon}       E-mail: Filipe.Bonjour@durham.ac.uk
  \bibitem[\ddagger]{rgre}      E-mail: R.A.W.Gregory@durham.ac.uk
  \bibitem[\S]{jste}            E-mail: john@amtp.cam.ac.uk
  \bibitem{VS}     A. Vilenkin and E. P. S. Shellard, {\it Cosmic Strings and
                   Other Topological Defects} (Cambridge University Press,
                   Cambridge, 1994).
  \bibitem{VA}     T. Vachaspati and A. Ach{\'u}carro, Phys. Rev. D {\bf 44},
                   3067 (1991); M. Hindmarsh, Phys. Rev. Lett. {\bf 68},
                   1263 (1991).
  \bibitem{V}      T. Vachaspati, Phys. Rev. Lett. {\bf 68}, 1977 (1992);
                   Phys. Rev. Lett. {\bf 69}, 216(E) (1992);
                   N. Manton, Phys. Rev. D {\bf 28}, 2019 (1983).
  \bibitem{T}      N. Turok, Phys. Rev. Lett. {\bf 63}, 2625 (1989).
  \bibitem{defect} R. Brandenberger, {\it Modern Cosmology and Structure
                   Formation} [astro-ph/9411049];
                   M. Hindmarsh and T. W. B. Kibble, Rep. Prog. Phys. {\bf 58},
                   477 (1995) [hep-ph/9411342].
  \bibitem{Int}    E. P. S. Shellard, Nucl. Phys. B {\bf 283}, 624 (1987);
                   K. J. M. Moriarty, E. Myers and C. Rebbi, Phys. Lett.
                   {\bf 207}, 411 (1988).
  \bibitem{SG}     A. Vilenkin, Phys. Rev. D {\bf 23}, 852 (1981);
                   J. R. Gott III, Astrophys. J. {\bf 288}, 422 (1985);
                   W. Hiscock, Phys. Rev. D {\bf 31}, 3288 (1985);
                   B. Linet, Gen. Relativ. Grav. {\bf 17,} 1109 (1985);
                   D. Garfinkle, Phys. Rev. D {\bf 32}, 1323 (1985);
                   R. Gregory, Phys. Rev. Lett. {\bf 59,} 740 (1987).
  \bibitem{Cri}    T. Vachaspati and A. Vilenkin, Phys. Rev. Lett. {\bf 67},
                   1057 (1991); D. N. Vollick, Phys. Rev. D {\bf 45}, 1884
                   (1992).
  \bibitem{MT}     K.-i. Maeda and N. Turok, Phys. Lett. {\bf202}B, 376 (1988);
                   R. Gregory, Phys. Lett. {\bf 206}B, 199 (1988).
  \bibitem{GG}     D. Garfinkle and R. Gregory, Phys. Rev. D {\bf 41}, 1889
                   (1990);
                   R. Gregory, D. Haws and D. Garfinkle, Phys. Rev. D {\bf 42},
                   343 (1990);
                   R. Gregory, Phys. Rev. D {\bf 43}, 520 (1991).
  \bibitem{SOA}    V. Silveira and M. D. Maia, Phys. Lett. {\bf 174}A, 280
                   (1993) [gr-qc/9303017];
                   P. Orland, Nucl. Phys. B {\bf 428}, 221 (1994)
                   [hep-th/9404140];
                   H. Arod\'z, Nucl. Phys. B {\bf 450}, 174 (1995)
                   [hep-th/9502018];
                   Nucl. Phys. B {\bf 450}, 189 (1995) [hep-th/9503001].
  \bibitem{NO}     H. B. Nielsen and P. Olesen, Nucl. Phys. B {\bf 61}, 45
                   (1973).
  \bibitem{F}      D. F\"orster, Nucl. Phys. B {\bf 81}, 84 (1974).
  \bibitem{CG}     B. Carter and R. Gregory, Phys. Rev. D {\bf 51}, 5839
                   (1995) [hep-th/9410095].
  \bibitem{A}      M. Anderson, Phys. Rev. D {\bf 51}, 2863 (1995).
  \bibitem{CGZ}    T. L. Curtright, G. I. Ghandour and C. K. Zachos, Phys.
                   Rev. D {\bf 34}, 3811 (1986);
                   B. Boisseau and P. S. Letelier, Phys. Rev. D {\bf 46}, 1721
                   (1992);
                   H. Arod\'z and A. L. Larsen, Phys. Rev. D {\bf 49}, 4154
                   (1994) [hep-th/9309089].
  \bibitem{Po}     A. Polyakov, Nucl. Phys. B {\bf 268}, 406 (1986).
  \bibitem{Kl}     H. Kleinert, Phys. Lett. {\bf 174}B, 335 (1986).
  \bibitem{Car}    B. Carter, Class. Quantum Grav. {\bf 11}, 2677 (1994).
  \bibitem{KT}     T. W. B. Kibble and N. Turok, Phys. Lett. {\bf 116}B, 141
                   (1982).
  \bibitem{TW}     D. Garfinkle and T. Vachaspati, Phys. Rev. D {\bf 42}, 1960
                   (1990).
  \bibitem{HB}     M. Sakellariadou, Phys. Rev. D {\bf 42}, 453 (1990);
                   Phys. Rev. D {\bf 43}, 4150(E) (1991).
  \bibitem{Spivak} M. Spivak, {\it Differential Geometry} (Publish or Perish,
                   Berkeley, CA, 1979).
  \bibitem{Carter} B. Carter, J. Geom. Phys. {\bf 8}, 53 (1992);
                   [hep-th/9705172].
  \bibitem{NAG}    Numerical Algorithms Group, {\it NAG Fortran Library Mark 16}
                   (Oxford, 1993).
  \bibitem{NRec}   W. H. Press, S. A. Teukolski, W. T. Vetterling and
                   B. P. Flannery, {\it Numerical Recipes in C: the Art of
                   Scientific Computing} (Cambridge University Press,
                   Cambridge, 1992).
\end{references}
\end{document}